Ю. И. Бельченко, Е. А. Гилев, З. К. Силагадзе

# МЕХАНИКА ЧАСТИЦ И ТЕЛ В ЗАДАЧАХ

НОВОСИБИРСК
2010

Сборник содержит задачи по механике и теории относительности, использующиеся при изучении курса общей физики на младших курсах физических факультетов университетов. Основу пособия составляют оригинальные задачи, предложенные преподавателями Новосибирского Государственного Университета – сотрудниками физических институтов Сибирского отделения РАН для семинаров, самостоятельной работы студентов, письменных работ. Задачи сгруппированы в соответствии с программой семинаров по механике, принятой в НГУ.

# ОГЛАВЛЕНИЕ

## 6. Динамика одномерного движения

## 7. Колебания

## 8. Движение в центральном поле

## 9. Движение твердого тела

## 10. Неинерциальные системы отсчета

# ПРЕДИСЛОВИЕ

Сборник содержит более 500 задач по курсу механики и теории относительности, предлагавшихся в течение многих лет студентам 1-го курса физического факультета Новосибирского государственного университета. Курс был разработан основателем и первым директором Института ядерной физики Сибирского отделения РАН, академиком Г.И. Будкером. Особенностью данного курса является изменение «классического» порядка изложения и введение современных представлений о свойствах пространства-времени и релятивистских законах движения с самого начала обучения, в рамках курса общей физики.

Основу сборника составляют оригинальные задачи, предложенные в разные годы коллективом преподавателей физики - научными сотрудниками физических институтов Сибирского отделения РАН и кафедр Новосибирского государственного университета для семинарских занятий, для контрольных заданий, для письменных и экзаменационных работ. Также включены классические задачи, взятые из учебников и учебных пособий, применяемых в курсе общей физики НГУ.

Сборник выпускался в виде учебных пособий в 1978, 1992, 2000 и 2006 гг, в 2008 г – в издательстве «Регулярная и хаотическая динамика» (М-Ижевск). Задачи сгруппированы по основным разделам курса в соответствии с программой семинаров, принятой в НГУ.

Авторы считают своим приятным долгом поблагодарить Б.В.Чирикова, Б.Н.Брейзмана, В.Г.Дудникова, П.А.Багрянского, В.Г.Соколова, Г.В.Федотовича и других физиков-преподавателей НГУ, внесших большой вклад в составление задач и подготовку предыдущих изданий сборника.

# ПРИНЯТЫЕ ОБОЗНАЧЕНИЯ

| | |
|---|---|
| $c$ | скорость света $= 3 \cdot 10^{8}$ м/с |
| Л-система | лабораторная система отсчета |
| Ц-система | система центра инерции |
| $x, y, z, t$ | координаты в лабораторной системе отсчета |
| $x', y', z', t'$ | координаты в движущейся системе отсчета |
| v, $V, u$ | скорость тела, системы отсчета |
| $\beta, \gamma, \gamma_{v}$ | релятивистские факторы |
| $\tau$ | собственное время |
| $T$ | интервал или период движения |
| $d, L$ | расстояние, длина |
| $a'$ | ускорение в сопутствующей системе отсчета |
| $r, R, \rho$ | радиус |
| $\alpha, \theta, \varphi$ | угол |
| $\Omega$ | телесный угол |
| $\omega, \Omega$ | угловая скорость |
| $v, f$ | частота колебаний |
| $m, M$ | масса |
| $E$ | энергия |
| $T$ | кинетическая энергия |
| $p, P$ | импульс |
| $L$ | момент импульса |
| $F$ | сила |
| $G$ | гравитационная постоянная, $G = 6,67 \cdot 10^{-11}$ H$\cdot$m$^2$кг$^{-2}$ |
| $q, Q, e$ | электрический заряд |
| $\mathcal{E}$ | напряженность электрического поля |
| $U$ | потенциал |
| $B$ | индукция магнитного поля |

# СПРАВОЧНЫЕ МАТЕРИАЛЫ

| | |
|---|---|
| 1 астрономическая единица (а. е.) | $1,5 \cdot 10^8$ км |
| 1 световой год (св. г) | $9,46 \cdot 10^{12}$ км |
| 1 парсек (пк) | $3,26$ св. г $= 3,1 \cdot 10^{13}$ км |
| Постоянная Хаббла | 71 (км/с)/Мпк |
| Гравитационная постоянная | $G = 6,67 \cdot 10^{-11}$ Н$\cdot$m$^2$кг$^{-2}$ |
| Радиус Солнца | $7 \cdot 10^5$ км |

## Земля

| | |
|---|---|
| Период обращения вокруг Солнца | 1 год $= \pi \cdot 10^7$ с |
| Средний радиус орбиты | $1,5 \cdot 10^8$ км $= 500$ св. с |
| Орбитальная скорость | 30 км/с |
| Угловая скорость вращения вокруг оси | $7,3 \cdot 10^{-5}$ рад/с |
| Угол наклона экватора к эклиптике | $23^{\circ}$ |

## Луна

| | |
|---|---|
| Орбитальная скорость | 1 км/с |
| Расстояние до Земли | $3,8 \cdot 10^5$ км |
| Период обращения | 27 сут |
| Масса | 1/81 массы Земли |
| Радиус | 1738 км |

## Планеты

| | Марс | Венера | Юпитер |
|---|---|---|---|
| Радиус орбиты, а. е. | 1,52 | 0,72 | 5,2 |
| Период обращения, год | 1,9 | 0,6 | 12 |
| Орбитальная скорость, км/с | 24 | 35 | 13 |

## Масса элементарных частиц

| | | | |
|---|---|---|---|
| Электрон $e^-$ | 511 кэВ | Протон $p$ | 938 МэВ |
| Позитрон $e^+$ | 511 кэВ | Нейтрон $n$ | 940 МэВ |
| Мюоны $\mu^-, \mu^+$ | 105 МэВ | $\phi$-мезон | 1 ГэВ |
| $\pi$-мезоны $\pi^-, \pi^+$ | 140 МэВ | $\Lambda$-гиперон | 1,1 ГэВ |
| $\pi^0$ | 135 МэВ | J/$\psi$-мезон | 3,1 ГэВ |
| Каоны $K^-, K^+, K^0$ | 500 МэВ | $B$-мезон | 5,28 ГэВ |

# 1. КИНЕМАТИКА

## 1.1. Пространство и время

**1.1.** Линейные размеры молекулы $O_2$ 0,3 нм, длина волны оранжевой линии криптона 605,8 нм, радиус Земли 6400 км, расстояние Земля – Луна 384 тыс. км, расстояние до $\alpha$-Центавра 4,2 св. года. Указать метод измерения этих величин и физические ограничения на точность измерения.

**1.2.** Скорость пули на начальном участке 715 м/с, скорость электрона в кинескопе $4 \cdot 10^7$ м/с, скорость света в вакууме $3 \cdot 10^8$ м/с. Указать способы измерения этих скоростей и физические ограничения на точность измерения.

**1.3.** Вычислить расстояние до звезды $\alpha$-Центавра по ее годичному параллаксу $\pi = 0{,}756''$ (в парсеках, световых годах и метрах).

**1.4.** Оценить радиус орбиты Венеры, если ее наибольшее угловое удаление от Солнца составляет $46^0$.

**1.5.** Определить максимальное время наблюдения Венеры после захода Солнца. Радиус орбиты Венеры 0,72 а. е.

**1.6.** Оценить максимальную продолжительность наблюдения полного лунного затмения. Видимый с Земли угловой размер Луны и Солнца имеет одинаковую величину $10^{-2}$ рад.

**1.7.** Оценить минимальную скорость движения лунной тени по поверхности Земли при солнечном затмении. Скорость Луны 1 км/с.

**1.8.** Как должна быть сориентирована спутниковая антенна НГУ для приема сигнала с геостационарного спутника? Найти максимальный угол наклона антенны к горизонту. Широта Новосибирска $55^0$.

**1.9.** Нарисуйте траекторию конца тени от вертикально стоящей палочки в солнечный день 22 июня в Новосибирске. Оцените долготу дня. Проследите эволюцию траектории со временем. Что будет на других широтах?

## 1.2. Системы координат. Скорость, ускорение

**1.10.** Зависимость скоростей двух автомобилей от времени задается следующими выражениями:

$$V_1 = \begin{cases} V\dfrac{t}{\tau} & t \le \tau \\ V\left(\dfrac{\tau}{t}\right)^2 & t > \tau \end{cases} \quad ; \qquad V_2 = \begin{cases} V\left(1-\cos\dfrac{\pi t}{\tau}\right) & 0 < t \le \tau \\ 2V & \tau < t \le \tau_1 \\ 2V\dfrac{\tau_2 - t}{\tau_2 - \tau_1} & \tau_1 < t \le \tau_2 \end{cases}$$

Найти зависимость ускорения и пройденного пути от времени. Нарисовать синхронные графики ускорения, скорости и пройденного пути.

**1.11.** График зависимости скорости объекта от времени имеет вид половинки окружности, занимающей на оси времени 20 с (в системе СИ), начальная скорость равна нулю. Какой путь пройден объектом за время движения? Нарисовать зависимость ускорения и пройденного пути от времени. Каким будет результат для половинки эллипса высотой 5 м/с?

**1.12.** Нарисовать синхронные графики зависимости от времени координаты $x$, скорости $\dot{x}$ и ускорения центра тяжести упругого шарика, подпрыгивающего в поле тяжести над упругой плитой без потерь энергии. Рассмотреть случай, когда деформации шарика существенны. За какое время шарик остановится, если при каждом ударе будет теряться 1 % энергии? Изобразите это движение на плоскости $x, \dot{x}$.

**1.13.** Упругий шарик подпрыгивает в поле тяжести над горизонтальной плитой, которая движется вниз с постоянной скоростью. Нарисовать синхронные графики зависимости от времени смещения $x$ шарика из начального положения, скорости $\dot{x}$ и ускорения $\ddot{x}$ в лабораторной системе отсчета. Изобразите это движение на плоскости $x$, $\dot{x}$.

**1.14.** Точка описывает фигуру Лиссажу по уравнениям

$$x = \cos\ t/\tau\ ,\quad y = 2\cos\ 2t/\tau\ ,$$

где $x$, $y$ заданы в м, $t$ – в с, $\tau = 1$ с. Определить скорость и ускорение точки, когда она пересекает ось OY. Нарисовать траекторию.

**1.15.** Движение точки выражается уравнениями

$$x = \cos\ t/\tau\ ,\quad y = 2\sin\ 2t/\tau\ ,$$

где $x$, $y$ заданы в м, $t$ – в с, $\tau = 1$ с. Определить зависимость проекций силы, действующей на точку, от координаты. Масса точки $10^{-3}$ кг.

**1.16.** Нарисовать зависимость от времени угла поворота, угловой скорости и углового ускорения антенны радиолокатора, следящего за самолетом, летящим по прямой с постоянной скоростью.

**1.17.** Зависимость скорости точки массой $m$ от времени имеет вид:

$$V_x = V_0 \sin\ \omega t + \varphi\ + V_1,\quad V_y = 0,\quad V_z = 0.$$

Нарисовать зависимость пути, пройденного точкой, от времени. Определить силу, действующую на точку.

**1.18.** Напряженность однородного электрического поля изменяется по закону $\mathcal{E} = \mathcal{E}_0 \cos\ \omega t + \varphi$. Нарисуйте траекторию движения электрона в таком поле, если в начальный момент $t = 0$ скорость движения электрона $V_0$ была направлена перпендикулярно полю.

**1.19.** Исследовать зависимость между напряжением $U(t)$ на пластинах осциллографа и смещением пятна на экране $y(t)$. Как изобразится на экране прямоугольный импульс напряжения? Найти уравнение кривой, на которой в момент времени $t_0$ находятся электроны луча.

**1.20.** На вертикальные пластины осциллографа подается напряжение $U_1(t)$, на горизонтальные – $U_2(t)$. Чувствительность осциллографа $\alpha_1$ и $\alpha_2$ В/м соответственно. Нарисовать траекторию светового пятна на экране, найти его скорость и ускорение в различные моменты времени для двух случаев:

1) $U_1 = at\cos\omega t, U_2 = at\sin\omega t;$

2) $U_1 = a\cos\omega t, U_2 = b\sin(\omega t + \varphi).$

**1.21.** Точка движется по закону

$$x = a \cdot ch\ \gamma t\ ,\quad y = a \cdot sh\ \gamma t\ ,$$

где *a, γ* — константы. Найти уравнение траектории в декартовых и полярных координатах, выразить скорость и ускорение точки как функцию ее радиус-вектора $r = \sqrt{x^2 + y^2}$.

**1.22.** Нарисовать траекторию точки, движущейся по закону

$$r = \frac{b}{t}, \ \ \varphi = \gamma t \ \ (b > 0).$$

Найти закон движения и уравнение траектории в декартовых координатах.

**1.23.** Получить выражение для компонент радиус-вектора, скорости и ускорения точки в цилиндрической системе координат.

**1.24.** Точка движется по закону

$$\rho = a \, e^{kt}, \ \ \varphi = kt.$$

Найти траекторию, скорость, ускорение и радиус кривизны траектории в зависимости от радиус-вектора точки.

**1.25.** Точка движется по закону $x = 2t, \ y = t^2 \ (x, \ y$ — в м, $t$ — в с). Определить радиус кривизны траектории в начале движения и через 2 с.

**1.26.** Установить связь между декартовыми, цилиндрическими и сферическими координатами. Записать выражения для дифференциала длины дуги, площади, объема и, используя их, вычислить прямым интегрированием длину окружности, площадь сферы, объем цилиндра, конуса, шара.

**1.27.** Четыре собаки преследуют друг друга, так что скорость догоняющей собаки $\vec{V}$ всегда направлена на убегающую собаку (см. рисунок). Через какое время собаки догонят друг друга, если сначала они находились в углах квадрата со стороной *a*? Какова траектория собак? Какими будут время и траектория для *N* собак?

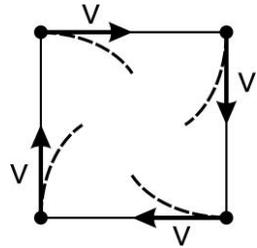

**1.28.** Корабль движется равномерно, сохраняя постоянный угол пеленга на маяк (угол между вектором скорости и направлением на маяк). Нарисовать возможные траектории корабля.

**1.29.** Заяц бежит по прямой линии со скоростью *и*. В начальный момент времени из положения, показанного на рисунке, его начинает преследовать собака со скоростью *V*. В ходе погони собака всегда бежит в направлении зайца. Через какое время собака настигнет зайца? Начальное расстояние между ними *L*.

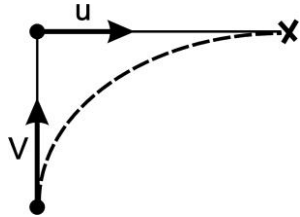

**1.30.** Имеется однородный шнур из взрывчатого вещества. Скорость распространения реакции взрыва вдоль шнура $V$, скорость распространения взрывной волны по воздуху $c$. Найти форму линии, по которой надо расположить шнур, чтобы волна от всех точек шнура пришла в заданную точку одновременно. Можно ли сделать то же самое для поверхности со взрывчаткой и получить сходящуюся сферическую волну с большой концентрацией энергии?

## 1.3. Векторы

**1.31**. Сферические координаты векторов

$$\vec{r}_1 = (r_1,\ \theta_1,\ \varphi_1)\ \text{и}\ \ \vec{r}_2 = (r_2,\ \theta_2,\ \varphi_2).$$

Определить угол между векторами $\vec{r}_1$ и $\vec{r}_2$.

**1.32**. Найдите кратчайшее расстояние при полете из Новосибирска ($\varphi = 83^0$ вост. долготы, $\theta = 55^0$ с. ш.) до Рио-де-Жанейро ($\varphi = 44^0$ зап. долготы, $\theta = 22^0$ ю. ш.).

**1.33.** Выразить орты сферической и цилиндрической систем координат через орты декартовой системы координат (и наоборот).

**1.34**. Футболист находится в 20 м от прямолинейной траектории мяча. Скорость мяча 10 м/с, футболиста 8 м/с. При каких начальных положениях мяча футболист сможет догнать его? В каких точках траектории мяча возможен перехват, если вначале мяч был в 25 м от футболиста?

**1.35.** Для двух кораблей, движущихся неизменными пересекающимися курсами, выразить расстояние наибольшего сближения и время до сближения через векторы скоростей и начальных положений.

**1.36.** Определить векторы скоростей и ускорений траков гусениц трактора, который движется по прямой дороге с ускорением $a$ при скорости $V$ (в системе дороги и в системе трактора).

**1.37.** Самолет облетел стороны треугольника с длинами $A$, $B$ и $C$ за время $t_1$, $t_2$, $t_3$ соответственно. Найти скорость ветра и самолета в случае, когда скорость ветра параллельна плоскости треугольника.

## 1.4. Системы отсчета

**1.38.** Снаряд, летящий на большой высоте со скоростью $V$, разрывается на осколки, которые в системе снаряда разлетаются в разные стороны с одинаковыми начальными скоростями *u*. Какое положение в пространстве они займут в момент времени *t*? Опишите движение осколков в системе координат, связанной с одним осколком.

**1.39.** Как изменяются импульс и кинетическая энергия системы частиц при преобразованиях Галилея? В какой системе отсчета кинетическая энергия частиц минимальна?

**1.40.** Какова кинетическая энергия гусеницы трактора в системе дороги и в системе трактора, если скорость трактора *V*?

**1.41.** Определите скорость поезда, если при приближении к неподвижному наблюдателю гудок поезда имел частоту в $\alpha$ раз большую, чем при удалении от наблюдателя.

**1.42.** Машинисты двух сближающихся поездов сигнализируют друг другу гудками. Определите скорость поездов, если частоты принимаемых машинистами сигналов превышают "собственную" частоту гудка в $\alpha$ и $\beta$ раз соответственно. Сигнальные устройства локомотивов одинаковы.

**1.43.** Какова угловая скорость вращения Луны с точки зрения наблюдателя, находящегося на поверхности Земли?

**1.44.** Найти траекторию и закон движения точки на циферблате в системе координат, связанной с концом минутной стрелки часов.

**1.45.** Найти траекторию и закон движения конца часовой стрелки часов в системе координат, связанной с концом минутной стрелки.

**1.46.** Найти траекторию и закон движения конца минутной стрелки часов в системе координат, связанной с концом часовой стрелки.

**1.47.** Нарисовать траекторию Марса в системе координат, связанной с центром Земли, начиная с противостояния. Период между двумя последовательными противостояниями Марса 780 земных суток. Расстояние Земля – Марс меняется от $0{,}55 \cdot 10^8$ км до $4 \cdot 10^8$ км.

**1.48.** Нарисовать траекторию Луны в системе координат, неподвижной относительно центра Солнца. Оценить диапазон ускорений центра Луны в этой системе координат.

**1.49.** Могут ли на траектории спутника Земли в системе, связанной с Солнцем, появиться участки с нулевой кривизной?

### 1.5. Кинематика вращения

**1.50.** Жесткий стержень *AB* движется в плоскости *XOY*, опираясь на окружность, центр которой находится в начале координат (см. рисунок).

Найти угловую скорость стержня, если его конец В движется вдоль оси *x* с постоянной скоростью *V*.

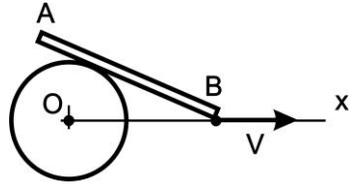

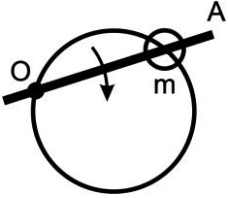

**1.51.** Стержень ОА (см. рисунок) равномерно вращается с угловой скоростью $\omega$ вокруг точки *О*, расположенной на окружности радиуса *R*. Определить скорость и ускорение колечка *m*, надетого на стержень и окружность.

**1.52.** Между двумя зубчатыми рейками зажата шестеренка радиусом *R* = 0,5 м. Ускорения реек $a_1 = 1,5$ м/с$^2$ и $a_2 = 2,5$ м/с$^2$ (см. рисунок). Найти поступательное и угловое ускорение шестеренки.

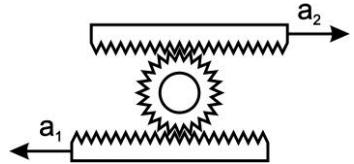

**1.53.** Стержень *AB* движется в плоскости *XOY*, опираясь своими концами на взаимно-перпендикулярные прямые *OX* и *OY*. Найти координаты мгновенного центра вращения в момент, когда угол *OAB* равен $60^0$.

**1.54.** По стенке дома затаскивают бревно длиной *L*, так что его верхний конец движется вертикально вверх с постоянной скоростью *V*, а нижний передвигается по земле. Найти угловую скорость и угловое ускорение точек бревна в различные моменты времени.

**1.55.** Найти мгновенный центр вращения и угловую скорость жесткого стержня, если известна величина и направление скорости одного конца и направление вектора скорости второго конца. Скорости концов стержня лежат в одной плоскости (см. рисунок).

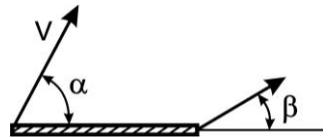

**1.56.** Конус, лежащий боковой поверхностью на горизонтальной плоскости, катится по ней без проскальзывания, так что его вершина неподвижна. Угол при вершине конуса $\alpha = 90^0$. Центр основания конуса движется равномерно и возвращается в начальное положение через 1 с. Найти вектор углового ускорения конуса.

**1.57.** В конической зубчатой передаче (см. рисунок) оси вращения шестерней неподвижны, $\omega_1 = 10$ об/мин; $\alpha = 30^0$, $\beta = 60^0$. Найти $\omega_2$.

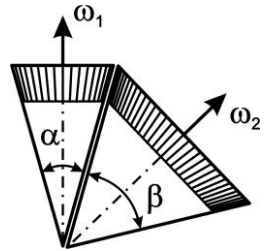

**1.58.** Определить закон движения и траекторию точки, находящейся на расстоянии $r$ от оси диска радиусом $R$, катящегося без проскальзывания по горизонтальной плоскости со скоростью $V$.

**1.59.** Найти угловую скорость, угловое ускорение и мгновенную ось вращения колеса автомобиля, когда он едет с постоянной по модулю скоростью $V$ по вогнутому мосту радиусом $R$. Радиус колеса $r$.

**1.60.** Найти мгновенный центр вращения эллиптического колеса, которое катится по вогнутому мосту с постоянной кривизной.

**1.61.** Найти угловую скорость, угловое ускорение и мгновенную ось вращения колеса трамвая при повороте. Колесо движется без проскальзывания с постоянной по модулю скоростью $V$ по рельсу с радиусом закругления $R$. Радиус колеса $r$.

# 2. РЕЛЯТИВИСТСКАЯ КИНЕМАТИКА

## 2.1. Скорость света

**2.1.** Две палочки, пересекающиеся под углом α, движутся поступательно со скоростями V перпендикулярно своей длине (см. рисунок). Найти скорость перемещения точки пересечения палочек. Может ли она превысить скорость света?

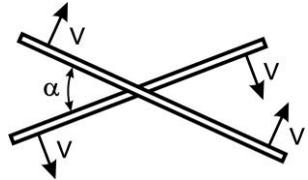

**2.2.** Фронт плоской волны падает под углом α, на плоскую поверхность AB (см. рисунок). Найти скорость перемещения точки F вдоль прямой AB. Можно ли считать эту скорость скоростью распространения некоторого сигнала вдоль прямой AB? Может ли она превысить скорость света?

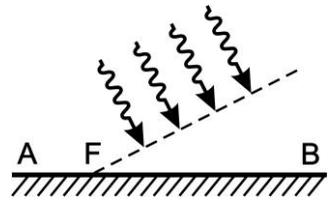

**2.3.** "Световой зайчик" от пульсара перемещается по поверхности Земли со скоростью $V =10^{20}$ м/с (угловая скорость вращения пульсара $\omega =$ 10 рад/с, расстояние до пульсара $10^{19}$ м). Можно ли скорость перемещения "зайчика" рассматривать как скорость распространения светового сигнала?

**2.4.** Космонавт находится в неосвещенном космическом корабле, летящем относительно Земли со скоростью, близкой к скорости света. На небольшом расстоянии от космонавта по ходу корабля расположено зеркало. Через какое время космонавт увидит свое изображение в зеркале после включения источника света, расположенного рядом с космонавтом?

**2.5.** На ракете, летящей со скоростью, близкой к скорости света, произошла вспышка света. С точки зрения ракеты область, занятая фотонами, представляет собой равномерно расширяющуюся сферу. Каким представляется волновой фронт от вспышки неподвижному наблюдателю?

**2.6.** За пять лет наблюдения с Земли светящийся объект, находящийся на расстоянии $10^5$ св. лет, совершил видимое угловое перемещение $10^{-4}$ рад, т. е. его кажущаяся скорость перемещения равна удвоенной скорости света. Найдите, под каким углом к линии наблюдения может двигаться объект, чтобы его реальная скорость была меньше скорости света. Какова минимально возможная скорость объекта?

**2.7.** Каким будет казаться земному наблюдателю время обращения спутника Ио вокруг Юпитера? Как меняется это время в течение года? Истинный период обращения Ио 42 часа.

## 2.2. События. Преобразования Лоренца

**2.8.** На концах стержня с собственной длиной $L_0$, движущегося со скоростью $V$, одновременно в системе стержня зажигаются две лампочки. Какая из них загорится раньше (и насколько) в Л-системе отсчета? Какую вспышку увидит раньше (и насколько) неподвижный наблюдатель, находящийся в точке O (см. рисунок)?

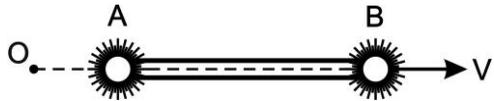

**2.9.** Масштаб $A'B'$ собственной длиной $L_0$ движется со скоростью $V$ вдоль такого же масштаба $AB$ (см. рисунок). Часы, находящиеся на концах масштабов в точках $A'$ и $B'$, $A$ и $B$, синхронизованы в своих системах отсчета. В момент совпадения точек $B'$ и $A$ часы $B'$ и $A$ показывали одинаковое время $t = 0$. Какое время показывают каждые часы в момент совпадения точек $A$ и $A'$; $B$ и $B'$?

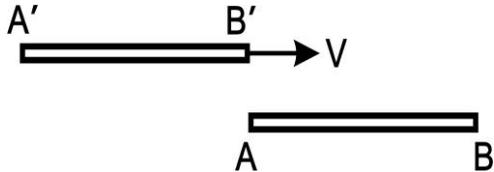

**2.10.** В точках 0 и $2L$ оси $x$ одновременно происходят вспышки света. В Л-системе отсчета фотоны этих вспышек «встречаются» в точках, равноудаленных от вспышек, т. е. в вертикальной плоскости, проходящей через точку $x = L$. Какой будет форма поверхности для точек встречи фотонов в системе отсчета, движущейся вдоль оси *x* с релятивистской скоростью *V*?

**2.11.** Квадратная (в собственной системе отсчета) платформа со стороной *L* движется вдоль своей диагонали со скоростью *V*. В углах платформы установлены зеркала. Отражаясь от них, по периметру платформы движется фотон. Найти период его движения в Л-системе отсчета.

**2.12.** Луч света движется через систему зеркал, расположенных в вершинах квадрата со стороной *L*. Найти время движения фотона через систему с точки зрения наблюдателей, движущихся со скоростью *V* = 0,8 *c* в направлениях, указанных на рисунке.

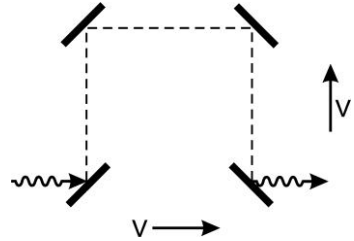

**2.13.** Космический корабль удаляется от Земли со скоростью *V*. Через время *T* после его старта с Земли посылают сигнал связи. Каково с точки зрения космонавтов расстояние между Землей и кораблем в момент получения сигнала?

**2.14.** Два космических корабля летят встречными курсами со скоростями *V* = 0,5 *c* каждый. При пролете мимо друг друга (в точке О рисунка) их часы были синхронизованы. Через час после встречи (по своим часам) один из кораблей посылает радиосигнал вдогонку другому. Какое время покажут часы, установленные на втором корабле, в момент приема посланного радиосигнала?

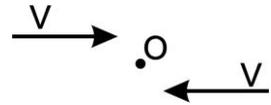

**2.15.** Космический корабль половину времени (по часам корабля) двигался с релятивистской скоростью $V_1$, а вторую половину – со скоростью $V_2$. Как далеко улетел корабль, если его путешествие по часам на Земле длилось время *T*?

**2.16.** Насколько должна отличаться от скорости света скорость образующихся на Солнце мюонов (время жизни $2 \cdot 10^{-6}$ с), чтобы от Солнца до Земли успевало долететь, не распавшись, 95 % частиц? Свет идет от Солнца до Земли 500 с.

**2.17.** В системе галактики фотон пролетает ее диаметр за время $T = 10^5$ лет. Сколько времени потребуется фотону на это путешествие в системе отсчета протона с релятивистским фактором $\gamma = 10^{10}$, летящего следом за фотоном? Как изменится результат, если протон летит навстречу фотону?

**2.18.** Из-за распада количество покоящихся нейтронов уменьшается экспоненциально с постоянной времени $10^3$ с. Какая доля нейтронов с релятивистским фактором $\gamma = 10^{10}$, стартовавших вместе с фотоном, "останется в живых" к моменту, когда фотон достигнет края галактики с размером $10^5$ световых лет? Рассмотреть задачу с точки зрения наблюдателя, движущегося вместе с нейтронами, и с точки зрения Л-системы отсчета.

**2.19.** Стеклянный брусок длиной $L$ движется со скоростью $V$ параллельно своей грани. Одна из сторон бруска, перпендикулярная к скорости, посеребрена. Сколько времени по часам неподвижного наблюдателя потребуется свету, летящему параллельно $V$, чтобы пройти сквозь брусок, отразиться от серебряной грани и выйти из бруска? Скорость света в неподвижном бруске $c/n$ ($n$ – показатель преломления).

**2.20.** Источник света находится на расстоянии $L$ от неподвижного зеркала. По пути от источника к зеркалу свет проходит через стеклянную пластинку толщиной $L_0$ с показателем преломления $n$, движущуюся вдоль луча со скоростью $V$. Найти время движения света от источника до зеркала и обратно.

**2.21.** На какое расстояние сдвинется вертикальный луч света после прохождения через стеклянный брусок, летящий горизонтально со скоростью $V \approx c$ (см. рисунок)? Какое время фотоны будут находиться в бруске? Показатель преломления стекла $n$, толщина бруска $d$.

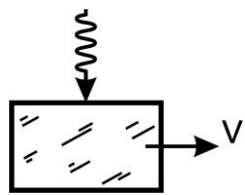

**2.22.** В вершине О прямоугольного треугольника АВО происходит вспышка света (см. рисунок). С какой скоростью и в каких направлениях может двигаться наблюдатель, чтобы в его системе отсчета свет достиг точки В раньше, чем точки А? $AO = AB = L$.

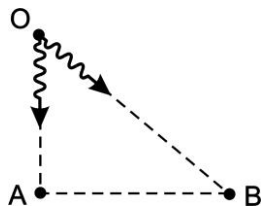

**2.23.** В релятивистскую реку со скоростью воды $V$ перпендикулярно линии берега из точки А на берегу бросили камень, упавший в воду на расстоянии $a$ от берега. За какое время волна от камня достигнет берега? Через какое время волна придет к точке А? Скорость волны в стоячей воде $u > V$.

**2.24.** Координаты двух событий в Л-системе отсчета $(\vec{r}_1, t_1)$ и $(\vec{r}_2, t_2)$. В какой системе отсчета эти события одновременны? В какой системе они одноместны? Сколько таких систем отсчета существует?

**2.25.** В центре сферы радиусом $R$ произошла вспышка света. С точки зрения наблюдателя, относительно которого сфера неподвижна, поверхность сферы освещается равномерно. Сколько времени сфера будет освещена с точки зрения наблюдателя, относительно которого сфера движется со скоростью $V$? Как соотносятся количества фотонов, поглощенных передней и задней частями сферы, с точки зрения этого наблюдателя?

**2.26.** Вдоль оси пенала длиной $L$, закрывающегося с торцов крышками А и В (см. рисунок *а*) со скоростью $V$ движется карандаш. Собственная длина карандаша $L_0$ удовлетворяет условию $L_0 > L > L_0/\gamma$ (где $\gamma$ – релятивистский фактор карандаша). Сначала крышка А пенала открыта, а крышка В закрыта. Когда карандаш влетает в пенал, крышка А закрывается, так что в течение некоторого времени карандаш находится в закрытом пенале (рис. *б*). Затем открывается крышка В и карандаш свободно вылетает из пенала. Опишите явление в системе отсчета карандаша. Что изменится, если у пенала не открывать крышку В?

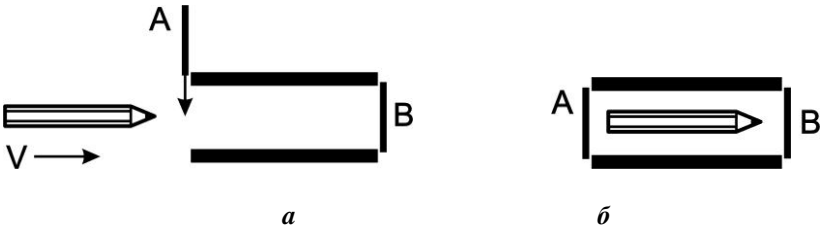

*а*                           *б*

**2.27.** Релятивистский трактор движется по полю с постоянной скоростью $V$. Сидящий в кабине тракторист насчитал на каждой половине гусеницы по $N$ траков. Сколько траков насчитает на верхней и нижней половинах гусеницы неподвижный наблюдатель?

**2.28.** Релятивистский танк движется по направлению к крепости со скоростью $V$. Он выпускает $n$ снарядов в секунду (по часам стрелка). Скорость снарядов относительно танка $u$. Сколько снарядов в секунду попадает в крепость (по часам гарнизона в крепости)?

**2.29.** В пучок релятивистских электронов, движущихся со скоростью $V$ и имеющих плотность частиц $n_-$, добавлено некоторое количество неподвижных однозарядных ионов с концентрацией $n_+ < n_-$, так что в Л-системе отсчета пучок заряжен отрицательно. Найти плотность частиц каждого сорта в системе отсчета, связанной с электронами. При каком условии суммарная плотность заряда в этой СО будет положительной, т. е. электроны будут стягиваться к оси пучка?

**2.30.** Показать, что площадь поперечного сечения, проведенного перпендикулярно направлению движения параллельного пучка света, является релятивистским инвариантом.

**2.31.** Между двумя линзами сформирован пучок света, имеющий круглое сечение радиусом $R$ и движущийся вертикально вниз (см. рисунок). Перпендикулярно пучку вдоль оси $x$ со скоростью $V$ движется диск такого же радиуса (в собственной системе отсчета). Плоскость диска перпендикулярна пучку. С точки зрения лабораторной системы отсчета диск испытывает лоренцево сокращение и не может перекрыть пучок света. Для наблюдателя на диске сокращается сечение пучка и должен наступить момент полного перекрытия света в фокусе второй линзы. Объясните парадокс.

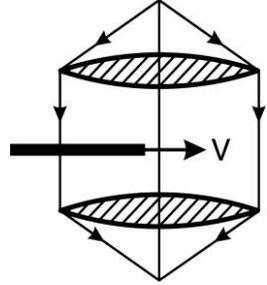

**2.32.** Параллельный пучок света падает на решетку, состоящую из брусков сечением $a$ x $d$ с расстоянием между брусками $b$ (см. рисунок). Какая часть падающего света сможет пройти через решетку, если ее двигать перпендикулярно пучку с релятивистской скоростью $V$?

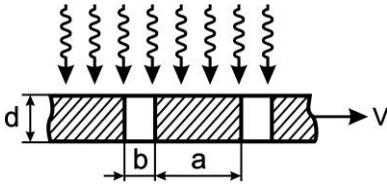

**2.33.** Что получится при моментальном фотографировании быстролетящих параллелепипеда, шара? Фотографирование производится в параллельных лучах света, падающих перпендикулярно фотопластинке.

### 2.3. Сложение скоростей

**2.34.** Две частицы движутся вдоль оси $x$ навстречу друг другу со скоростями $V_1$ и $V_2$. Найти скорость сближения частиц в Л-системе отсчета и сравнить ее с относительной скоростью частиц.

**2.35.** Один из двух одинаковых стержней покоится, а другой движется вдоль него со скоростью $V$. В какой системе отсчета длины стержней будут равными?

**2.36.** Массивная плита, движущаяся перпендикулярно своей плоскости со скоростью $c/2$, налетает на легкую неподвижную частицу. Найти скорость частицы после упругого столкновения с плитой.

**2.37.** В ускорителе на встречных пучках ток электронного пучка равен току позитронного, а плотность электронов в системе отсчета позитронного пучка в *N* раз больше, чем плотность позитронов. Какова скорость пучков? Лабораторная геометрия пучков одинакова.

**2.38.** Системы отсчета $S_1$ и $S_2$ движутся вдоль оси *x* со скоростями $V_1$ и $V_2$ относительно системы *S*. На часах, покоящихся в системе $S_1$, секундная стрелка совершает один оборот. Сколько времени длится этот оборот с точки зрения системы $S_2$?

**2.39.** Космический корабль, летящий к Земле со скоростью $V = 0,6\ c$, посылает к Земле ракету связи, движущуюся относительно корабля со скоростью $u = 0,8\ c$. Через какое время по часам корабля и Земли ракета встретится с Землей, если в момент ее старта корабль находился на расстоянии 5 св. лет от Земли (в СО Земли)?

**2.40.** Найти скорость распространения света относительно покоящегося наблюдателя, если луч движется в среде с показателем преломления *n*, которая движется относительно наблюдателя со скоростью *V* (опыт Физо).

**2.41.** Релятивистский эскалатор, движущийся со скоростью $V = c/3$, имеет *N* ступенек. Пассажир, опаздывающий на поезд, сбегает по эскалатору со скоростью $u = c/2$ ( относительно эскалатора), не пропуская ни одной ступеньки. Сколько шагов сделает пассажир по эскалатору?

**2.42.** Ширина релятивистской реки $2L = 2$ км, скорость воды у берега равна нулю и нарастает к середине реки. Найти скорость течения в центре реки, полагая, что в пределах узкой полоски шириной $\Delta x$ в системе отсчета ближайшего к берегу края полоски скорость воды нарастает линейно по закону $\Delta V = \alpha \Delta x$, где $\alpha = 3 \cdot 10^5\ \text{с}^{-1}$.

**2.43.** Найти скорость одинаковых частиц в системе центра масс, если в Л- системе отсчета их скорости параллельны, но не равны по величине.

**2.44.** Найти скорость двух одинаковых частиц в системе центра масс, если в Л-системе отсчета их скорости $\vec{V}_1$ и $\vec{V}_2$.

**2.45.** Из двух точек, разделенных расстоянием *L*, одновременно вылетают две частицы с перпендикулярными друг другу одинаковыми по величине скоростями *V* (см. рисунок). Найти минимальное расстояние между частицами:

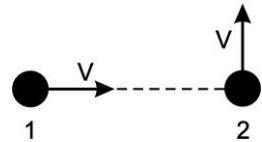

а) в Л-системе отсчета;
б) в системе одной из частиц.

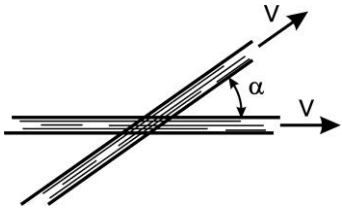

**2.46.** Два пучка частиц, летящих со скоростью *V*, пересекаются под углом α (см. рисунок). Найти плотность частиц в зоне пересечения пучков в системе отсчета одного из пучков. Плотность каждого из пучков в Л-системе равна *n*.

**2.47.** Цилиндрический пенал сечением *S* и собственной длиной $L_0$ движется с релятивистской скоростью *V* навстречу неподвижному облаку пыли плотностью *ρ*. Когда первые пылинки прилипают к дну пенала B (см. рисунок), по его стенкам начинает распространяться упругая волна, движущаяся относительно пенала со скоростью *u*. Когда волна достигает переднего конца пенала, его крышка A закрывается. Сколько вещества окажется внутри пенала?

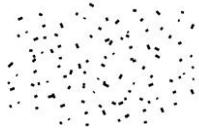
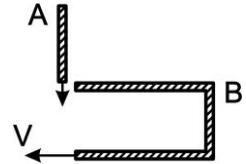

## 2.4. Направление движения. Преобразование углов

**2.48.** На берегу релятивистской реки, текущей со скоростью *V*, стоит мальчик и кидает в воду камни. Под каким углом к линии берега он должен кидать камни, чтобы до него доходили волны от места падения? Скорость волн в стоячей воде равна *u*.

**2.49.** Катер, имеющий относительно воды скорость *u*, движется из пункта A в пункт B, находящийся строго напротив на другом берегу релятивистской реки шириной *L*, текущей со скоростью *V*. За какое время он пересечёт реку?

**2.50.** Скорость течения релятивистской реки $V = 0,6\,c$, скорость релятивистского пловца относительно воды $u' = 0,8\,c$. За какое минимальное время в системе отсчета берега пловец может переплыть реку, если ее ширина $10^3$ м? Каким будет минимальное время в системе отсчета пловца?

**2.51.** Прозрачная пластинка с показателем преломления *n*, толщиной *d* движется параллельно своей плоскости с релятивистской скоростью *V*. В точке A, расположенной снаружи от пластинки, произошла вспышка света. В какую точку *X* Л-системы отсчета, расположенную на другой стороне пластинки (см. рисунок), свет дойдет раньше всего? Свет в пластинке распространяется со скоростью *c/n*.

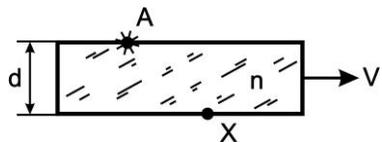

**2.52.** На противоположных берегах релятивистской реки со скоростью течения *V* и шириной *L* находятся пункты А и В. Пункт А расположен ниже по течению пункта В на расстоянии *S* в системе берега. Под каким углом к берегу должен плыть релятивистский пловец, чтобы попасть из В в А? Скорость пловца относительно воды *u'*. Рассмотреть задачу в системе отсчета берега и в системе отсчета воды.

**2.53.** Луч света падает вертикально к поверхности стеклянного прямоугольного бруска (см. рисунок), движущегося вдоль горизонтальной оси со скоростью *V*. Найти угол к вертикали, под которым луч выйдет через боковой торец бруска после преломления, и условие, при котором свет выйдет из бруска. Показатель преломления стекла *n*. Закон преломления света на границе двух сред в системе бруска $n_1 \sin \alpha_1 = n_2 \sin \alpha_2$.

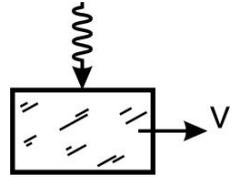

**2.54.** При быстром движении наблюдателя относительно небосвода в передней полусфере насчитывается в *N* раз больше звезд, чем в задней. Определите скорость этого движения, если для неподвижного наблюдателя звезды распределены по небу изотропно.

**2.55.** Звезда, летящая со скоростью *V* порядка *c*, "сбрасывает" часть своего вещества, причем в системе отсчета звезды осколки разлетаются изотропно со скоростью *u'*. В какой телесный угол по ходу движения звезды будет лететь половина сбрасываемого вещества (с точки зрения неподвижного наблюдателя)? Какая часть сбрасываемого вещества будет лететь вперед по ходу движения звезды?

**2.56.** Найти изменение направления скорости частицы при переходе в движущуюся систему отсчета.

**2.57.** Космический корабль начинает двигаться, сохраняя постоянным угол $\alpha < \pi/2$ между вектором скорости корабля *V* и направлением на радиомаяк (в своей системе отсчета). При каких значениях скорости корабль будет приближаться к маяку? За какое время корабль совершит полный оборот вокруг маяка, если начальное удаление от маяка было *R*?

**2.58.** Найти форму видимой кривой, которую описывает на небосводе очень удаленная звезда вследствие годичной аберрации.

**2.59.** Стержень длины *L* падает на пол со скоростью *u* (ускорением пренебречь), так что его концы достигают пола одновременно. Какой будет длина стержня и под каким углом к полу он будет расположен с точки зрения наблюдателя, движущегося по полу со скоростью *V* параллельно плоскости падения стержня?

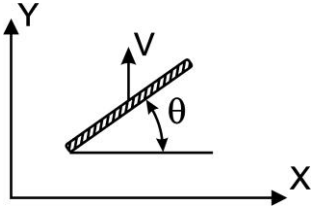

**2.60.** Стержень движется со скоростью $V = 0,6\ c$ вдоль оси *y*, причем его ось наклонена под углом $\theta = 45^\circ$ к оси *x* (см. рисунок). Каким будет угол наклона стержня для наблюдателя, движущегося со скоростью $u = 0,8\ c$ вдоль оси *x*?

**2.61.** Два стержня с собственной длиной $L_0$, ориентированные перпендикулярно друг другу, движутся с одинаковыми по величине скоростями *V* вдоль своей длины (см. рисунок). Какова длина одного из стержней в системе отсчета другого стержня? Каким будет результат в случае, если стержни движутся перпендикулярно своей длине?

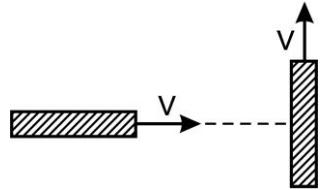

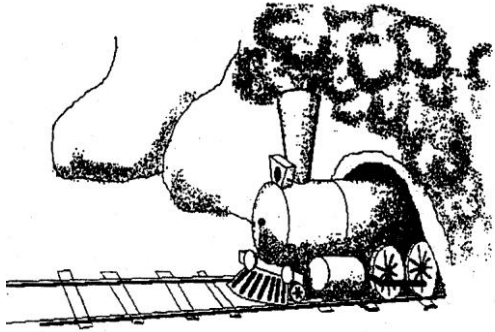

# 3. ЭНЕРГИЯ–ИМПУЛЬС

## 3.1. Масса, энергия и импульс релятивистских частиц

**3.1.** Вычислить скорость и импульс электронов (в $\dfrac{эВ}{c}$), если их кинетическая энергия равна:

      а) 5 кэВ  (трубка осциллографа);

      б) 3 МэВ  (электростатический  ускоритель Ван-де-Граафа);

      в) 50 ГэВ  (Стэнфордский линейный ускоритель).

**3.2.** Заряженные π-мезоны с импульсом 54 МэВ/с пролетают в среднем расстояние $L$ = 3 м (от момента рождения до распада). Найти собственное время жизни π-мезонов. Масса π-мезона 140 МэВ.

**3.3.** Энергия каждого из протонов в накопительном кольце равна $2 \cdot 10^{13}$ эВ, средний ток пучка 70 мА. Какая мощность будет выделяться, если этот пучок направить на поглощающую мишень? Сколько энергии выделится в мишени, если дорожка накопителя имеет длину 87 км? Пучок заполняет камеру равномерно по ее длине.

**3.4.** Найти давление, производимое пучком электронов с кинетической энергией 3 МэВ и током 1 А, сфокусированным на поглощающей мишени в пятно площадью 1 мм$^2$ .

**3.5.** Оценить давление света, излучаемого настольной лампой, на лежащий под ней лист бумаги.

**3.6.** Оцените энергию и давление реликтового излучения, считая, что оно равномерно заполняет Вселенную размером $1{,}3 \cdot 10^{10}$ св. лет. "Тем-пература" квантов реликтового излучения $3 \cdot 10^{-3}$ эВ, плотность энергии $6 \cdot 10^{-14}$ Дж/м$^3$.

**3.7.** Плотность мощности солнечного излучения на орбите Земли имеет величину $W = 1,4$ кВт/м$^2$. Насколько меняется масса Солнца в единицу времени за счет излучения света?

**3.8.** Давлением лазерного луча зеркало удерживается в поле тяжести. Какова масса зеркала, если мощность лазера 200 кВт?

### 3.2. Преобразования энергии-импульса. Эффект Доплера

**3.9.** Лазер испускает импульс света длительностью $T$ и полной энергией $E$, который отражается от идеального зеркала, приближающегося к лазеру со скоростью $V$. Каковы энергия и длительность светового импульса после отражения? Свет падает по нормали к поверхности зеркала.

**3.10.** Фотон движется между двумя параллельными зеркалами, одно из которых неподвижно, а другое приближается к первому со скоростью $V \ll c$. Фотон движется по нормали к зеркалам. Как изменится энергия фотона при уменьшении расстояния между зеркалами в два раза? Изобразить движение фотона на фазовой плоскости. Как изменяется площадь, ограниченная его фазовой траекторией?

**3.11.** На космическом корабле, удаляющемся от Земли со скоростью $V = c/2$, вышла из строя энергетическая установка. Чтобы обеспечить корабль энергией, с Земли посылают лазерный луч. Какова должна быть мощность лазера, если на борту корабля потребляется мощность $N$?

**3.12.** На космическом корабле, удаляющемся от Земли со скоростью $V$ порядка $c$, включают передатчик мощностью $N$. Какова максимальная мощность сигнала, принимаемого на Земле антенной площадью $\pi r^2$? В момент включения передатчика корабль находился на расстоянии $L \gg r$.

**3.13.** Пучок электронов со средней энергией частиц 50 ГэВ и относительным разбросом энергий $\pm 1$ % движется вдоль оси $x$ Л-системы отсчета. Какова максимальная кинетическая энергия электронов в сопровождающей пучок системе отсчета?

**3.14.** Электроны в пучке имеют среднюю энергию 100 кэВ и энергетический разброс $\pm 1$ эВ. Каков разброс энергий электронов ($T_e$) в сопровождающей пучок системе отсчета?

**3.15.** Найти скорость центра инерции для системы частиц с импульсами $\vec{p}_k$ и энергиями $E_k$.

**3.16.** Докажите, что суммарная кинетическая энергия ансамбля невзаимодействующих частиц имеет минимальную величину в системе центра масс.

**3.17.** В 1929 г. Э. Хаббл установил, что галактики разбегаются с относительными скоростями $V = H \cdot r$, пропорциональными расстояниям $r$ между ними, где константа Хаббла $H = 71$ (км/с)/Мпарсек. Найти расстояние до галактики, если известно, что спектр приходящего от нее излучения при 10 %-м сжатии шкалы длин волн совпадает со спектром излучения нашей галактики.

**3.18.** Определите скорость и рабочую длину волны радиомаяка НЛО, если при его приближении к Земле принимаемая частота радиосигнала была $v$, а при удалении – $v/4$.

**3.19.** Космический корабль, летящий со скоростью $V_1$, посылает сигнал частотой $\omega_1$ и после его отражения от летящего навстречу другого космического корабля принимает сигнал частотой $\omega_2$ (частоты даны в системе отсчета первого корабля). Найти скорость второго корабля.

**3.20.** Источник света, движущийся горизонтально со скоростью $V = 0{,}6\ c$ на расстоянии $R$ от пола, приближается к стенке высотой $H = 0{,}5\ R$ (см. рисунок). В какую точку на полу за стенкой свет придет быстрее всего? Свет какой частоты увидит наблюдатель в этой точке в момент появления света? Частота света в системе отсчета источника $\omega_0$.

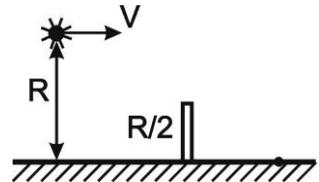

**3.21.** Релятивистский трактор движется со скоростью $V$ навстречу неподвижному локатору. Определить максимальную и минимальную частоту сигналов, возвращающихся к локатору после однократного отражения от гусениц трактора. Частота сигнала локатора $v_0$.

**3.22.** Источник света с частотой $v_0$ движется со скоростью $V$ в неподвижной среде с показателем преломления $n$ мимо неподвижного наблюдателя. Какую частоту будет регистрировать наблюдатель при приближении источника света и при его удалении?

**3.23.** Две подводные лодки движутся навстречу друг другу с релятивистскими скоростями $V_1$ и $V_2$ относительно воды. Радары лодок имеют одинаковую собственную частоту $v_0$, скорость сигнала в воде равна $c/n$. Какова частота сигналов, принимаемых на лодках?

**3.24.** Для определения скорости космического объекта, пролетающего мимо Земли, его зондируют лазерным лучом с частотой фотонов $\nu_0$. Определите скорость объекта по частоте $\nu_H$ фотонов, "вернувшихся" к наблюдателю, и по углу $\theta$ между направлением движения объекта и лучом света (см. рисунок).

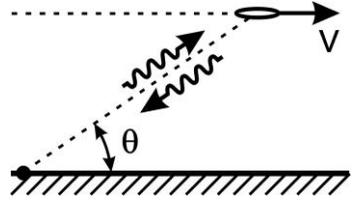

**3.25.** Ракета летит со скоростью $V$ через скопление, состоящее из $N$ одинаковых, равномерно распределённых звёзд. Сколько звёзд, по мнению экипажа, уменьшило длину волны своего излучения более чем в два раза по сравнению с результатами наблюдений из неподвижного корабля?

**3.26.** Луч света, параллельный оси $x$, отражается от массивного зеркала, движущегося вдоль оси $x$. Какова скорость и угол наклона зеркала, если частота света при отражении уменьшается в 2 раза, а отраженный луч движется по нормали к оси $x$?

**3.27.** Найти изменение направления движения фотона при отражении от массивного зеркала, движущегося со скоростью $V$ перпендикулярно своей плоскости.

**3.28.** Уголковый отражатель сделан из двух массивных зеркал, соединенных под углом $90^0$ (см. рисунок). Луч света, падающий под любым углом на неподвижный отражатель, меняет направление движения на противоположное. Какими будут частота и угол отражения света, если отражатель движется со скоростью $V$ вдоль оси $x$, а падающий луч – под углом $\theta$ к оси $x$? Частота падающего света $\nu_0$.

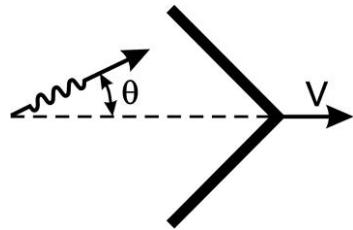

**3.29.** Система зеркал, закрепленных под углом $90^0$ друг к другу, движется со скоростью $V = c/2$ перпендикулярно лучу света частотой $\nu$ (см. рисунок). Какова энергия фотонов, испытавших двойное отражение от зеркал? Под каким углом будет двигаться свет после двойного отражения от зеркал?

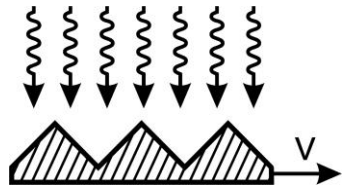

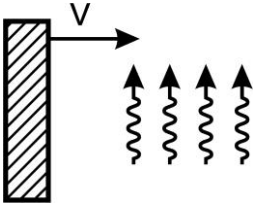

**3.30.** Идеальное зеркало движется со скоростью $V = 0{,}6\ c$ через световой поток с плотностью энергии $W = 10^{-2}$ Дж/м$^3$. Плоскость зеркала параллельна направлению движения фотонов (см. рисунок). Найти давление света на зеркало.

**3.31.** Идеальное двухстороннее зеркало ускоряется лучом лазера. Какой будет установившаяся скорость зеркала, если с противоположной стороны его тормозит луч второго лазера в четыре раза меньшей мощности? Свет падает по нормали к поверхности зеркала.

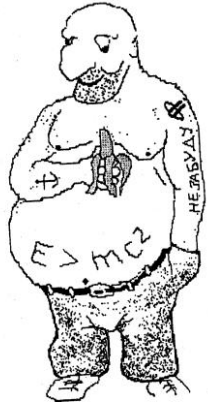

# 4. ЗАКОНЫ СОХРАНЕНИЯ ЭНЕРГИИ И ИМПУЛЬСА

## 4.1. Распад частиц

**4.1.** Пучки мюонов получают при распаде заряженных π-мезонов на мюон и нейтрино. На каком удалении от мишени, генерирующей поток π-мезонов импульсом 10 $\frac{ГэВ}{с}$, примесь π-мезонов к потоку мюонов становится меньше 50 % ? Масса заряженного π-мезона 140 МэВ, время жизни $2{,}6 \cdot 10^{-8}$ с.

**4.2.** Найти энергию $\pi^0$-мезонов, распадающихся по схеме $\pi^0 \rightarrow \gamma + \gamma$, если счетчик, расположенный по направлению их движения, регистрирует γ-кванты распада с энергией 270 МэВ. Масса $\pi^0$-мезона 135 МэВ.

**4.3.** Найти энергию $\pi^0$-мезонов, распадающихся по схеме $\pi^0 \rightarrow \gamma + \gamma$, если энергия летящих назад γ-квантов распада 10 МэВ. Масса $\pi^0$-мезона 135 МэВ.

**4.4.** Релятивистский снаряд скоростью $V = 0{,}6\ c$ и массой $M$ разрывается на лету на $N$ осколков с нулевой массой. Какова максимально возможная энергия одного из осколков, если в системе отсчета снаряда распад изотропен? Сравните со случаем, когда распад в системе снаряда не изотропен.

**4.5.** Энергия возбуждения ядра $Fe^{57}$ равна 14,4 кэВ. На сколько отличается от энергии возбуждения энергия фотона, испущенного незакрепленным возбужденным ядром? Какова скорость ядра после испускания фотона?

**4.6.** Летящий π-мезон распадается на мюон и нейтрино: $\pi^+ \to \mu^+ + \nu_\mu$. Найти энергию π-мезонов, если известно, что максимальная энергия рождающихся при распаде нейтрино в $\alpha = 100$ раз больше минимальной. Масса $\pi^+$-мезона 140 МэВ. Массу нейтрино считать равной нулю.

**4.7.** При распаде π-мезонов на мюоны и нейтрино $\pi^+ \to \mu^+ + \nu_\mu$ максимальные импульсы регистрируемых мюонов в $\alpha = 10$ раз больше минимальных. Найти энергию π-мезонов. $M_{\pi+} = 140$ МэВ, $m_\mu = 105$ МэВ.

**4.8.** Покоящийся ϕ-мезон распадается на две частицы: $\phi \to X + \gamma$. X-частица, в свою очередь, распадается на два γ-кванта: $X \to 2\gamma$. Известно, что отношение максимальной и минимальной энергий этих двух квантов равно $\alpha = 3{,}47$. Найти массу X-частицы. Масса ϕ-мезона 1020 МэВ.

**4.9.** $\pi^0$-мезон распадается на два γ-кванта. Найти энергию распадающихся $\pi^0$-мезонов, если счетчик, расположенный под углом $45^0$ к направлению их движения, регистрирует γ-кванты с энергией 27 МэВ.

**4.10.** Найти массу J/ψ-мезона, распавшегося на электрон и позитрон с одинаковыми энергиями 3,1 ГэВ и углом разлета $60^0$.

**4.11.** Какова энергия $\pi^+$-мезонов, распадающихся на лету по схеме $\pi^+ \to \mu^+ + \nu$, если энергия образовавшегося мюона $E_\mu = 300$ МэВ, а угол разлета образовавшихся частиц $\theta = 60^0$. $M_{\pi+} = 140$ МэВ, $M_\mu = 105$ МэВ.

**4.12.** $\pi^0$-мезон распадается на лету на два γ-кванта, углы вылета которых составляют соответственно $\theta_1$ и $\theta_2$ с начальным направлением движения $\pi^0$-мезона. Найти энергию $\pi^0$-мезона, если его масса равна $M$.

**4.13.** Определить угловой размер конуса, в который попадает половина γ-квантов, образующихся при распаде $\pi^0$-мезонов с энергией $E = 10^{10}$ эВ. Масса $\pi^0$-мезона 135 МэВ. Найти минимальную и максимальную энергию квантов, попавших в этот конус.

**4.14.** Определить угловой размер конуса, в который полетят осколки разрывного снаряда, летящего со скоростью 0,99 $c$. Масса снаряда $M$. Снаряд разлетается на $N$ осколков массой $M/3N$ каждый.

**4.15.** Пучок $\pi^0$-мезонов, летящих со скоростью $V = 0{,}5$ $c$, распадается по схеме $\pi^0 \to \gamma + \gamma$. Во сколько раз суммарная энергия образовавшихся фотонов, летящих в переднюю полусферу (по направлению движения пионов), больше, чем суммарная энергия фотонов, летящих назад? В системе отсчета $\pi^0$-мезонов распад изотропен.

**4.16.** Гиперон распадается на нейтрон и $\pi$-мезон. При какой минимальной энергии пучка гиперонов будут отсутствовать $\pi$-мезоны, летящие навстречу пучку? Масса гиперона 1,2 ГэВ, нейтрона 940 МэВ, $\pi$-мезона 140 МэВ.

**4.17.** $\pi$-мезон с энергией 500 МэВ распадается на мюон и нейтрино. Найти максимальный угол вылета мюона по отношению к направлению движения $\pi$-мезона. Масса $\pi$-мезона 140 МэВ, мюона 105 МэВ.

**4.18.** Найти связь между углами вылета $\theta_1$ и $\theta_2$ в Л-системе при распаде на две частицы.

**4.19.** Частица, летящая со скоростью $V$, распадается на две частицы. Найти связь между углами вылета и энергией образовавшихся частиц.

**4.20.** Летящий каон распадается по схеме $K^+ \rightarrow \pi^+ + \pi^+ + \pi^-$. На какое максимальное расстояние от линии движения каона успевают удалиться $\pi$-мезоны за время своей жизни? При какой минимальной энергии пучка каонов будут отсутствовать $\pi$-мезоны, летящие навстречу пучку? Собственное время жизни $\pi^+$-мезона $\tau = 2{,}6 \cdot 10^{-8}$ с, его масса $M_{\pi^+} = 140$ МэВ, масса каона $M_K = 494$ МэВ.

**4.21.** Определить интервал значений, которые может принимать угол между направлениями вылета распадных частиц равной массы в Л-системе при распаде на две частицы.

**4.22.** При распаде каонов с энергией 510 МэВ по схеме $K^0 \rightarrow \pi^+ + \pi^-$ минимальный угол разлета $\pi$-мезонов равен $150^0$. Определите массу каона.

**4.23.** Каон с энергией $10^{10}$ эВ и массой $5 \cdot 10^8$ эВ распадается по схеме $K^0 \rightarrow \pi^+ + \pi^-$. Найти максимальный угол разлета $\pi$-мезонов. Какова при этом относительная скорость $\pi$-мезонов?

**4.24.** Метастабильная молекула, обладающая внутренней энергией $E$ и кинетической энергией $T$, распадается на две частицы массы $m_1$ и $m_2$. Определить диапазон возможных направлений движения частиц в лабораторной системе.

**4.25.** Неподвижный $\phi$-мезон распадается на фотон и X-частицу, которая, в свою очередь, распадается на два фотона $\phi \rightarrow X + \gamma \rightarrow 2\gamma + \gamma$. Минимальный угол разлета двух фотонов при распаде X-частицы $\theta_{min} = 113°$. Найти массу X-частицы, если масса $\phi$-мезона 1020 МэВ.

**4.26.** Каоны, летящие в пучке со скоростью $V = 2/3\ c$, распадаются на мюон и нейтрино. Найти отношение числа мюонов, вылетающих в одинаковые малые телесные углы вперед и назад относительно направления движения каона. Масса каона 494 МэВ, мюона – 105 МэВ, массу нейтрино считать равной нулю. Каким будет это отношение для нейтрино?

**4.27.** Место распада монохроматического пучка $\pi^0$-мезонов с энергией 1 ГэВ окружено счетчиками, регистрирующими энергии всех $\gamma$-квантов распада $\pi^0 \to \gamma + \gamma$. Нарисовать регистрируемое распределение $\gamma$-квантов по энергиям.

**4.28.** Найти распределение распадных частиц по энергиям в Л-системе, если в Ц-системе угловое распределение имеет вид $dN \simeq \sin^2\theta\, d\Omega$, где $\theta$ – угол между скоростью $V$ первичной частицы и направлением вылета распадной частицы в Ц-системе. Скорость распадных частиц в Ц-системе равна $V_0$.

**4.29.** Какая доля $\gamma$-квантов, рождающихся при распаде летящих параллельно $\pi^0$-мезонов с энергией 10 ГэВ на два $\gamma$-кванта, имеет энергии, отличающиеся от максимальной не более чем на 10 %? Определить угловой раствор конуса, в который попадают $\gamma$-кванты.

**4.30.** Для нейтрино $\nu$, образующихся при распаде $\pi$-мезонов с энергией 6 ГэВ ($\pi^+ \to \mu^+ + \nu$), определить энергетический спектр, максимальную и среднюю энергию, а также угловое распределение в лабораторной системе отсчета, если известно, что в системе отсчета $\pi$-мезона распад изотропен. Массу нейтрино считать равной нулю.

**4.31.** Определить наибольшую кинетическую энергию, которую может получить частица с массой $m_1$, образуемая при распаде неподвижной частицы массой $M$ на несколько осколков с суммарной массой $m$.

**4.32.** Найти максимальную энергию электрона, возникающего при распаде неподвижного мюона: $\mu^- \to e + \nu + \tilde{\nu}$.

**4.33.** При какой максимальной энергии мюона электрон, получающийся при распаде $\mu^- \to e + \nu + \tilde{\nu}$, будет покоиться в лабораторной системе отсчета?

**4.34.** Неподвижный атом позитрония (связанная кулоновскими силами электрон-позитронная пара) аннигилирует в три $\gamma$-кванта. Углы разлёта γ-квантов в двух разных парах составили $120^0$ и $150^0$. Найти энергии всех трёх γ-квантов. Энергией связи электрона и позитрона в атоме позитрония пренебречь.

## 4.2. Неупругие столкновения. Пороги рождения частиц

**4.35.** При столкновении встречных пучков электронов и позитронов с энергиями 300 МэВ возможна реакция $e^+ + e^- \to 2\pi$. Каждый из образующихся π-мезонов затем распадается на два γ-кванта. Найти максимальную энергию образующихся γ-квантов.

**4.36.** Счетчик, предназначенный для регистрации π-мезонов, рождающихся при столкновении электронов и позитронов в установке со встречными пучками (по реакции $e^+ + e^- \to \pi^+ + \pi^-$), удален на 6 м от места встречи. Во сколько раз нужно увеличить показания счетчика, чтобы скомпенсировать распад π-мезонов за время пролета до счетчика, если энергия частиц в каждом пучке 280 МэВ? Масса π-мезона 140 МэВ, время жизни $2,6 \cdot 10^{-8}$ с.

**4.37.** Доказать, что излучение и поглощение света свободным электроном в вакууме невозможно. Возможна ли однофотонная аннигиляция электрон-позитронной пары по схеме $e^+ + e^- \to \gamma$?

**4.38.** Найти угол симметричного разлета фотонов $\alpha$ (см. рисунок), получающихся при аннигиляции покоящегося электрона с движущимся позитроном.

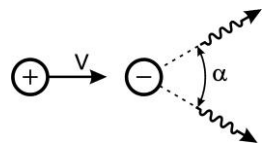

**4.39.** До какой энергии нужно ускорить позитроны, чтобы на мишени из покоящихся электронов проявлялись эффекты, возможные на установке со встречными электрон-позитронными пучками при энергии частиц в каждом пучке 20 ГэВ?

**4.40.** Антипротон при столкновении с неподвижным протоном образует электрон-позитронную пару: $p^+ + p^- \to e^+ + e^-$. Какова энергия образовавшегося электрона, если он движется перпендикулярно направлению первоначального движения антипротона?

**4.41.** При какой энергии протонов становится возможным рождение $J/\psi$-мезонов с массой $3,1 \cdot 10^9$ эВ на мишени из неподвижных протонов по схеме $p^+ + p^+ \to p^+ + p^+ + J/\psi$? При какой энергии электронов и позитронов наблюдается рождение *J/ψ*-мезона в экспериментах на встречных электрон-позитронных пучках?

**4.42.** Найти пороговую энергию рождения электрон-позитронной пары при столкновении γ-квантов с разными энергиями: $\gamma + \gamma \to e^+ + e^-$. Энергия "холодного" γ-кванта 1 эВ.

**4.43.** Какова суммарная кинетическая энергия нуклонов, образующихся в реакции $\gamma + d \to n + p$ при пороговой для этой реакции энергии γ-квантов? Энергия связи дейтрона 2 МэВ.

**4.44.** При облучении неподвижной водородной мишени $\pi^-$-мезонами идет реакция $\pi^- + p \to \gamma + n$. Найти максимальную энергию γ-кванта, если кинетическая энергия π-мезона 300 МэВ, масса 140 МэВ.

**4.45.** Определить скорости протона и π-мезона (масса 135 МэВ), образующихся при столкновении фотона с первоначально покоившимся протоном $\gamma + p \to \pi + p$. Фотон имел пороговую для этой реакции энергию.

**4.46.** Движущееся возбужденное ядро массой *M* и энергией возбуждения *E* испускает γ-квант. При какой скорости ядра испущенный γ-квант может возбудить аналогичное неподвижное ядро, находящееся в основном состоянии, до энергии возбуждения *E*?

**4.47.** При каких энергиях π-мезона γ-квант, полученный при распаде $\pi \to \gamma + \gamma$ и летящий назад, может при столкновении с тяжелым ядром образовать электрон-позитронную пару?

**4.48.** Найти максимальную и минимальную энергию, уносимую нейтрино в реакции взаимодействия электрона с энергией *E* и неподвижного протона: $e + p \to n + \nu$. Массу нейтрино считать равной нулю.

**4.49.** Найти порог рождения пары протон-антипротон при столкновении электрона и позитрона, скорости которых в лабораторной системе равны по величине и перпендикулярны.

**4.50.** Электрон и позитрон с энергией 5 ГэВ каждый сталкиваются и рождают протон и антипротон: $e^+ + e^- \to e^+ + e^- + p^+ + p^-$. При каких значениях угла между импульсами электрона и позитрона это возможно?

**4.51.** Две частицы массами $m_1$ и $m_2$, летящие со скоростями $\vec{V}_1$ и $\vec{V}_2$, слипаются в одну. Найти массу и скорость образовавшейся частицы.

**4.52.** Электрон с кинетической энергией 1 МэВ аннигилирует при столкновении с покоящимся позитроном. Один из двух образовавшихся при аннигиляции γ-квантов вылетает перпендикулярно направлению движения электрона. Найти направление вылета и энергию второго γ-кванта.

**4.53.** Предполагается, что протоны космических лучей из удаленных источников не доходят до Земли, если их энергия больше некоторой, так что они поглощаются при столкновениях с фотонами реликтового излучения в пороговой реакции $p + \gamma \to \Delta^+$. Найти величину пороговой энергии, если масса $\Delta^+$ резонанса равна 1232 МэВ. Масса протона 938 МэВ, энергия фотона реликтового излучения $E_\gamma = 3 \cdot 10^{-4}$ эВ.

**4.54.** На встречных электрон-позитронных пучках происходит реакция $e^+ + e^- \to \pi^+ + \pi^- + \gamma$. При какой минимальной энергии встречных пучков энергия образующегося γ-кванта будет больше 50 МэВ? Масса π-мезона 140 МэВ.

**4.55.** При столкновении встречных электрон-позитронных пучков с неодинаковыми энергиями образуется ϒ-мезон массой 10,58 ГэВ, который распадается на два B-мезона: $e^+ + e^- \to \Upsilon \to 2\,B$. При какой минимальной разнице энергий встречных пучков оба B-мезона будут двигаться в одну сторону? Какой при этом будет энергия ϒ-мезона? Масса B-мезона 5,28 ГэВ.

**4.56.** Во встречных электрон-позитронных столкновениях наблюдается процесс рассеяния с излучением фотона: $e^+ + e^- \to e^+ + e^- + \gamma$. Электрон и позитрон разлетаются под углами $\theta_1$ и $\theta_2$ по отношению к линии движения фотона. Найти энергию фотона, если энергии частиц встречных пучков $E$. Все электроны и позитроны ультрарелятивистские.

**4.57.** В электрон-позитронных столкновениях с различной энергией встречных пучков $E_2 \neq E_1$ происходит процесс $e^+ + e^- \to \psi' \to \tau^+ + \tau^-$. Масса Ψ′-мезона 3,7 ГэВ. Масса τ-лептона 1,8 ГэВ, время жизни $\tau = 3 \cdot 10^{-3}$ с. Каким должно быть отношение энергий встречных пучков $\kappa = E_2 / E_1$, чтобы в случае симметричного разлета τ-лептоны успевали отлететь от места рождения в среднем на расстояние $L = 5 \cdot 10^{-5}$ м?

**4.58.** Пучок фотонов с энергией $E_0$ сталкивается со встречным пучком фотонов различных энергий. В результате одного из столкновений рождается частица, которая распадается на два фотона. Найти массу этой частицы, если один из образовавшихся при распаде фотонов вылетел с энергией $E_1$ под углом $90^0$ к направлению движения исходного пучка фотонов. Какова была энергия встречного фотона?

**4.59.** Найдите мощность фотонных двигателей ракеты, необходимую для поддержания ее равномерного движения со скоростью 0,99 $c$ через межзвездный водородный газ плотностью $10^6$ м$^{-3}$. Поперечное сечение ракеты 40 м$^2$, столкновения с водородом считать неупругими.

**4.60.** Ракета с фотонным двигателем движется в облаке "космической пыли". Вся пыль, встречающаяся на пути ракеты, улавливается и используется как топливо при аннигиляции с таким же количеством антивещества в фотонных двигателях. Найти предельную скорость ракеты.

### 4.3. Упругие столкновения

**4.61.** Частица массой $m_1$ налетает на покоившуюся частицу массой $m_2$. При каком условии налетающая частица после упругого лобового столкновения будет двигаться назад? Рассмотреть нерелятивистский и релятивистский случаи.

**4.62.** Какую максимальную энергию может передать частица массой $m$ с кинетической энергией $T$ при упругом столкновении с первоначально покоившейся частицей массой $M$? Рассмотреть нерелятивистский и релятивистский случаи.

**4.63.** Найти минимальную кинетическую энергию, которая может остаться у частицы массой $m_1$ после упругого столкновения с покоящейся частицей массой $m_2$ в а) нерелятивистском и б) релятивистском случаях.

**4.64.** Какую энергию получает покоившийся протон после рассеяния электрона с энергией 1 ГэВ на угол $10^{-3}$ рад?

**4.65.** Нерелятивистская частица массой $m_1$, скоростью $V$ упруго сталкивается с неподвижной частицей массой $m_2$. Выразить скорость каждой из частиц после столкновения через угол ее рассеяния в Л-системе.

**4.66.** На какой максимальный угол может отклониться $\alpha$-частица с кинетической энергией 1 МэВ при упругом столкновении с первоначаль-но покоившимся нейтроном? Масса $\alpha$-частицы в 4 раза больше массы нейтрона.

**4.67.** Частица массой $m$ испытывает упругое столкновение с неподвижной частицей массой $M$. Выразить угол рассеяния в Ц-системе через угол рассеяния первой частицы в Л-системе (нерелятивистский случай).

**4.68.** Поток нерелятивистских моноэнергетических нейтронов в результате однократных упругих столкновений рассеивается на первоначально покоившихся протонах. Считая рассеяние изотропным в Ц-системе, найти долю нейтронов, рассеявшихся в Л-системе на угол, больший $60^0$.

**4.69.** Моноэнергетический пучок нейтронов с кинетической энергией 1 МэВ упруго рассеивается на первоначально покоившихся ядрах $He^4$. Найти функцию распределения рассеянных нейтронов по энергиям в Л-системе, если в Ц-системе рассеяние изотропно. Определить среднюю энергию однократно рассеянных нейтронов.

**4.70.** Определить энергию электрона, если при лобовом столкновении с фотоном реликтового излучения с энергией $3 \cdot 10^{-4}$ эВ он отдает половину своей энергии.

**4.71.** Электрон испытывает лобовое столкновение с неподвижным протоном и передает ему половину своей энергии. Найти начальную энергию электрона.

**4.72.** Какую максимальную энергию могут приобрести фотоны с энергией 2 эВ при рассеянии на встречном пучке электронов с энергией $10^{10}$ эВ?

**4.73.** Ультрарелятивистский электрон сталкивается с неподвижным электроном. Найти его энергию, если известно, что после рассеяния оба электрона имеют ультрарелятивистские энергии и летят под малыми углами $\theta_1$ и $\theta_2$ к линии первоначального движения электрона. Найти энергии электронов после рассеяния.

**4.74.** Фотон с энергией $E$ сталкивается с покоящимся электроном. Найти энергию фотона, рассеянного на угол $\theta$. Насколько изменится длина волны фотона при таком столкновении?

**4.75.** Фотон с энергией 10 эВ рассеивается на угол $90^\circ$ на электроне, летящем навстречу. Найти энергию рассеянного фотона, если кинетическая энергия электрона была: а) 100 эВ; б) 10 ГэВ.

**4.76.** На какой максимальный угол может отклониться релятивистский протон при столкновении с первоначально неподвижным электроном?

**4.77**. Определить минимальный угол разлета релятивистских частиц после упругого столкновения, если массы обеих частиц одинаковы, а одна из частиц до удара покоилась.

**4.78.** Два фотона с энергиями $E_1$ и $E_2$ сталкиваются под углом α друг к другу и рассеиваются с изменением своих энергий. Чему равна минимальная и максимальная энергия у рассеянных фотонов? Найдите минимальный угол между рассеянными фотонами.

**4.79.** В системе отсчета, движущейся со скоростью $V$ порядка $c$ относительно Л-системы, два электрона летели навстречу друг другу с энергией $E'$ каждый и рассеялись на малый угол $\theta'$. Найти изменение энергии электронов в Л-системе отсчета.

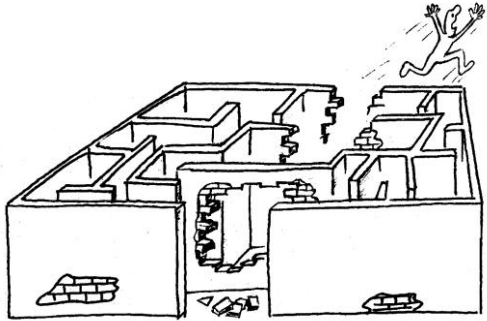

# 5. ПРОСТЕЙШАЯ РЕЛЯТИВИСТСКАЯ ДИНАМИКА

## 5.1. Движение в магнитном поле. Сила Лоренца

**5.1.** Во сколько раз различаются кинетические энергии протона и электрона с одинаковыми импульсами 200 $\dfrac{\text{МэВ}}{c}$ ? При какой напряженности магнитного поля радиус траекторий таких частиц равен 1 м?

**5.2.** Оценить кинетические энергии протонов, электронов и мюонов, проходящих по одинаковым траекториям радиусом 0.3 м через поворотный магнит с полем 1 Т.

**5.3.** Оценить напряженность магнитного поля, отклоняющего электроны энергией 10 кэВ на угол $60^0$ в телевизионной трубке. Отклоняющая катушка создает магнитное поле на участке трубки длиной 10 см.

**5.4.** Оценить максимальное смещение электронного луча на экране при повороте осциллографа вокруг вертикальной оси. Энергия электронов 10 кэВ, длина трубки 50 см, магнитное поле Земли 0,5 Гс.

**5.5.** Каким должен быть радиус у кольцевого ускорителя с поворотным магнитным полем 1 Т, предназначенного для ускорения протонов до энергии $10^{11}$ эВ? При какой величине магнитного поля в такой ускоритель следует инжектировать протоны с энергией $3 \cdot 10^8$ эВ?

**5.6.** Какой минимальный радиус должен иметь электрон-позитронный ускоритель со встречными пучками, чтобы на нем можно было наблюдать рождение Z-бозонов по схеме $e^+ + e^- \to Z$? Масса Z-бозона 90 ГэВ. Магнитное поле на дорожке ускорителя 1 Т.

**5.7.** При столкновении ультрарелятивистских электрон-позитронных пучков из двух одинаковых по размерам кольцевых накопителей наблюдается рождение ϒ-частицы ($e^+ + e^- \to \Upsilon$). Определите скорость ϒ-частицы, если напряжённости магнитного поля на дорожках накопителей отличаются в три раза. Масса ϒ-частицы 10 ГэВ.

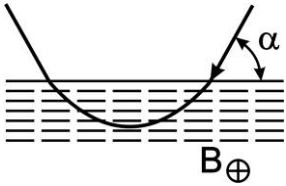

**5.8.** Показать, что угол падения равен углу отражения для электрона, влетающего под углом $\alpha$ в полупространство с неоднородным магнитным полем (см. рисунок). Поле перпендикулярно плоскости рисунка и изменяется лишь в направлении, перпендикулярном границе.

**5.9.** Мюоны влетают под углом $\alpha$ в полупространство с однородным поперечным магнитным полем величиной $B = 10^{-2}$ Т. Какая часть мюонов выйдет из области магнитного поля, не распавшись? Масса мюона 105 МэВ, собственное время жизни $2 \cdot 10^{-6}$ с.

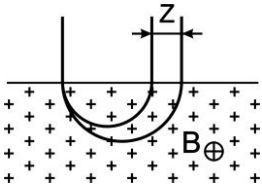

**5.10.** Пучок протонов со средней энергией $E = 2$ ГэВ и энергоразбросом $\pm 2$ % инжектируется в полупространство с однородным поперечным магнитным полем $B = 1$ Т (см. рисунок). Каким будет ширина пучка на выходе из магнитного поля?

**5.11.** Пучок протонов и электронов с одинаковыми энергиями 2 ГэВ проходит через магнитный фильтр, который состоит из двух участков длины 0,2 м с однородным и перпендикулярным пучку магнитным полем величиной 1 Т разной полярности (см. рисунок). Найти расстояние между пучками протонов и электронов после прохождения фильтра.

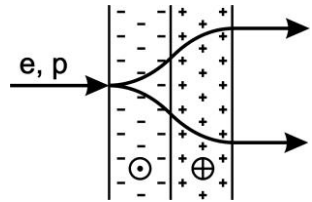

**5.12.** Мюоны движутся по круговой траектории радиусом 10 м в поперечном магнитном поле величиной 1 Т. За какое время мюонный ток уменьшится в 20 раз? Собственное время жизни мюона $2 \cdot 10^{-6}$ с, масса 105 МэВ.

**5.13.** В ускорителе типа микротрон электрон ускоряется из состояния покоя, проходя узкий ускоряющий зазор резонатора, где получает добавку энергии $\Delta E = mc^2 = 0{,}511$ МэВ. После этого он совершает оборот во внешнем поперечном магнитном поле и снова попадает в ускоряющий зазор (см. рисунок). Найти время ускорения электрона до энергии 511 МэВ в магнитном поле величиной $B = 2$ Т.

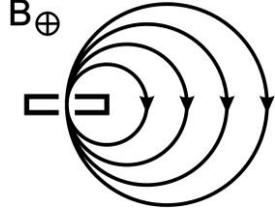

**5.14.** Каким будет радиус орбиты электронов в микротроне (см. рисунок к задаче 5.13) при ускорении до энергии 50 МэВ, если за один пролет резонатора электрон получает энергию 0,5 МэВ? Во сколько раз отличаются максимальный и минимальный радиусы орбит в этом случае? Напряженность магнитного поля 1 Т.

**5.15.** Гамма-квант с энергией 10 МэВ сталкивается с тяжелым ядром, рождая электрон-позитронную пару. Треки образовавшихся электрона и позитрона в детекторе с магнитным полем представляют собой окружности с радиусами 4 и 6 см. Найти кинетические энергии образовавшихся частиц.

**5.16.** В пузырьковую камеру, которая находится в поперечном однородном магнитном поле, влетает поток каонов $K^-$. Трек каона в камере имеет вид дуги окружности (см. рисунок). При распаде каона по схеме

$K^- \rightarrow \mu^- + \bar{\nu}_\mu$ отношение максимального радиуса трека образующегося мюона к минимальному радиусу равно $N = 3$ (для треков, лежащих в плоскости, перпендикулярной полю). Найти отношение масс каона и мюона $M_K/M_\mu$. Время жизни каона $1{,}2 \cdot 10^{-8}$ с, средняя длина трека каона до распада в камере 1,85 м.

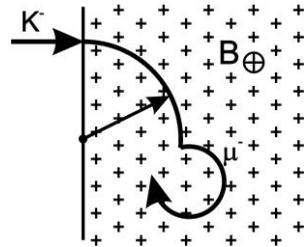

**5.17.** В пузырьковой камере с магнитным полем $B = 2$ Т фиксируется распад $\Lambda_0 \rightarrow \pi + p$. Доказать, что точка пересечения треков протона и $\pi$-мезона лежит на линии движения $\Lambda_0$-гиперона. Найти максимальное расстояние $L$ от точки распада до точки пересечения треков. Треки частиц лежат в плоскости, перпендикулярной магнитному полю. $M_p = 938$ МэВ, $M_\pi = 140$ МэВ, $M_\Lambda = 1110$ МэВ.

**5.18.** Релятивистская частица массой $m$, зарядом $q$ и энергией $3mc^2$ движется в поперечном магнитном поле $B$. За какое время ее энергия уменьшится до $2mc^2$, если на частицу действует сила трения $F = -\alpha V$? Сколько оборотов при этом совершит частица?

**5.19.** Определить траекторию движения релятивистской заряженной частицы в поперечном постоянном магнитном поле $B$ при наличии силы вязкого трения $F = -\alpha V$.

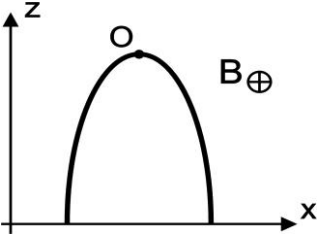

**5.20.** Электрон со скоростью $V$ влетает по нормали в область поперечного магнитного поля, увеличивающегося в направлении $z$ по закону $B = B_0 (1 + z/a)$. Определите радиус кривизны траектории электрона в верхней точке траектории О (см. рисунок).

**5.21.** Релятивистский электрон движется по окружности радиуса $r_0$ в аксиально-симметричном магнитном поле $B_z = B_0 (r_0 / r)^n$. Как изменится движение электрона, если его энергию увеличить на $\delta E$?

## 5.2. Движение в электрическом поле. Релятивистская ракета

**5.22.** Электрон влетает в тормозящее постоянное однородное электрическое поле напряженностью $\mathcal{E}$ с начальной скоростью $V_0 \parallel \mathcal{E}$. Через какое время электрон вернется в начальную точку? Какой путь он пройдет за это время?

**5.23.** Поток мюонов влетает по нормали в область тормозящего электрического поля напряженностью $\mathcal{E} = 10^6$ В/м. При какой энергии мюонов из области поля после "отражения" выйдет более 50 % частиц? Масса мюона 105 МэВ, время жизни $2 \cdot 10^{-6}$ с.

**5.24.** Ракета удаляется от Земли с постоянным ускорением $a'$ в сопутствующей системе отсчета. Через время $T$ после старта ей вдогонку посылается сигнал связи. За какое время он догонит ракету? При каком $T$ сигнал уже не сможет догнать ракету?

**5.25.** Сколько времени займет по земным часам и по часам космонавтов путешествие с удалением на 20 световых лет и возвращением обратно, совершаемое с постоянным по модулю ускорением, равным $a' = 9{,}8$ м/с$^2$ (1 св. г /г$^2$) в сопутствующей системе отсчета? Изобразите это путешествие в пространстве событий.

**5.26.** Какой должна быть длина линейного ускорителя со средней напряженностью электрического поля $10^7$ В/м, предназначенного для ускорения мюонов до энергии $10^{10}$ эВ? За какое время мюон с нулевой начальной энергией ускорится до $10^{10}$ эВ? Какая доля впущенных мюонов ускорится до конечной энергии?

**5.27.** До какой энергии ускорится мюон из состояния покоя в однородном электрическом поле напряженностью $\mathcal{E} = 10^7$ В/м за время жизни в собственной системе отсчета $2 \cdot 10^{-6}$ с? Масса мюона 105 МэВ. Какой будет казаться длина такого ускорителя с точки зрения ускоряемых мюонов, если длиной считать суммарную длину участков ускорителя, измеренную из сопутствующих мюону систем отсчета?

**5.28.** Протон и электрон ускоряются из состояния покоя навстречу друг другу однородным электрическим полем напряженностью $\mathcal{E} = 10^7$ В/м. При каком минимальном расстоянии $L$ между точками старта протона и электрона возможно образование пионов при столкновении в реакции $p + e^- \rightarrow p + e^- + \pi^- + \pi^+$? Каким будет это расстояние, если протон стартует раньше электрона на время $T = 10^{-7}$ с? $M_\pi = 140$ МэВ.

**5.29.** Две стартовавших одновременно ракеты движутся одна за другой по прямолинейной траектории с одинаковым ускорением $g$ (в сопутствующих системах отсчета). В момент старта, когда расстояние между ракетами было равно $d = 0,5$ св. года, они послали друг другу радиосигналы частотой $\omega$. Какова частота сигналов, принятых на каждой из ракет?

**5.30.** Частица массой $m$ влетает с начальной скоростью $V_0$ в область, где на нее действует постоянная сила $\vec{F} \perp \vec{V}_0$. К каким значениям стремятся составляющие импульса $p_\text{п}, p_\perp$ и скорости $V_\text{п}, V_\perp$, направленные соответственно вдоль и поперек начальной скорости?

**5.31.** Электрон с кинетической энергией 1 МэВ влетает в тормозящее однородное электрическое поле напряженностью $\mathcal{E} = 10^6$ В/м под углом $\alpha = 45^0$ (см. рисунок). Какова высота траектории и минимальная скорость электрона? Найти расстояние между точками влета и вылета электрона и время его пролета через конденсатор.

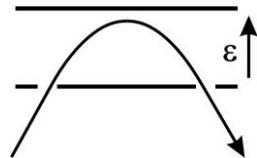

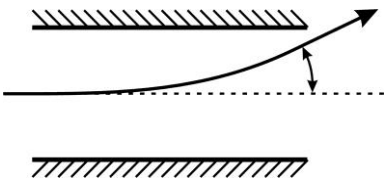

**5.32.** При пролете через электростатический отклоняющий конденсатор (см. рисунок) протон с кинетической энергией $10^6$ эВ отклоняется на угол 0,1 радиана. На какой угол отклонится в этом конденсаторе электрон с такой же кинетической энергией?

**5.33.** Точечный радиоактивный источник, расположенный в однородном электрическом поле с напряженностью $10^6$ В/м, испускает изотропный поток электронов с энергией $10^4$ эВ. Найти диаметр электронного потока на экране, перпендикулярном электрическому полю и находящемся на расстоянии 1 м от источника.

**5.34.** На наклонной плоскости с углом наклона α в поле тяжести лежит монета. Монете сообщается релятивистская скорость $V$ вдоль горизонтальной оси (см. рисунок). Найти установившуюся скорость движения монеты $u$, если коэффициент трения μ = tg α не зависит от скорости.

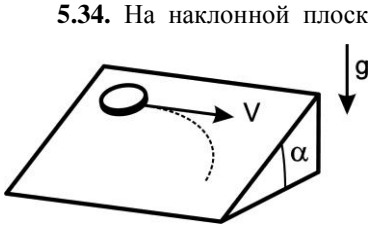

**5.35.** Идеальное зеркало массой 1 кг ускоряется лучом лазера, расположенного на Земле. Какой должна быть мощность лазера, чтобы ускорить зеркало до скорости $0{,}8\,c$ за один год?

**5.36.** Космический корабль движется по круговой траектории с постоянной скоростью $V = 0{,}6\,c$, причем центростремительное ускорение в сопутствующей системе отсчета равно $a'_n = g$. За какое время по собственным часам корабль совершит полный оборот?

**5.37.** Релятивистский электрон движется в однородных электрическом $\mathcal{E}$ и магнитном $B$ полях $(\vec{\mathcal{E}} \parallel \vec{B})$. Во сколько раз продолжительность $N + 1$ периода обращения электрона по винтовой траектории больше продолжительности предыдущего периода? В начальный момент времени импульс электрона был перпендикулярен полю.

**5.38.** Источник излучения свободно движется со скоростью $V$. В собственной системе отсчёта источник начинает изотропно излучать мощность $N$. Найти силу, действующую на источник в лабораторной системе отсчёта, и изменение его импульса и скорости за время $t$.

**5.39.** Заряженная частица, движущаяся с ускорением, излучает. Интенсивность излучения в сопутствующей системе отсчёта $I \sim a^2$, где $a$ — ускорение частицы, причем импульс излучения в сопутствующей системе отсчета равен нулю. Во сколько раз отличаются скорости потерь энергии на излучение у релятивистских протона и электрона в Л-системе, если они движутся:

а) в продольном постоянном однородном электрическом поле,

б) в магнитном поле по окружности одинакового радиуса?

**5.40.** Какую скорость приобретет космический корабль с фотонным двигателем, когда его масса уменьшится вдвое? Корабль ускоряется от нулевой начальной скорости. Считать, что масса топлива много больше массы ракеты, а к.п.д. двигателя равен 100 %.

**5.41.** Какую часть топлива израсходует ракета с фотонным двигателем при доускорении от скорости $0,9\,c$ до $0,99\,c$?

**5.42.** Первую половину пути ракета ускоряется с постоянным в сопутствующей системе отсчета ускорением $a' = 9,8$ м/с$^2$ (1 св. г /г$^2$), а вторую половину тормозится с таким же ускорением. Сколько горючего потребуется для полета на расстояние 10 св. лет? Конечная масса ракеты $M_k$, скорость истечения газов относительно ракеты $u$.

# 6. ДИНАМИКА ОДНОМЕРНОГО ДВИЖЕНИЯ

## 6.1. Законы движения. Фазовая плоскость

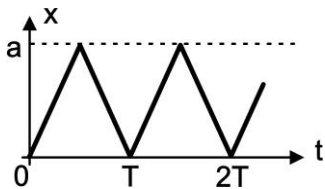

**6.1.** Найти зависимость силы, действующей на частицу, от координаты, если частица движется по закону, показанному на рисунке. Нарисовать зависимость потенциала от координаты. Изобразить движение на фазовой плоскости.

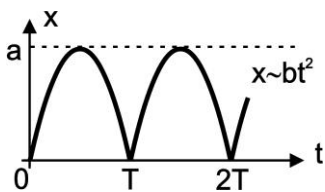

**6.2.** Найти зависимость силы, действующей на частицу, от координаты, если частица движется по закону, показанному на рисунке. Нарисовать зависимость потенциала от координаты. Изобразить движение на фазовой плоскости.

**6.3.** Исследовать движение заряженной частицы в однородном электрическом поле с синусоидальной зависимостью от времени. Нарисовать синхронные графики зависимости смещения, скорости и ускорения частицы от времени, а также траектории на фазовой плоскости при различных начальных условиях.

**6.4.** Найти зависимость силы, действующей на частицу массы $m$, от координаты, если закон движения частицы имеет вид $x(t) = a\sin\omega t$. Нарисовать зависимость потенциала от координаты. Изобразить движение на фазовой плоскости.

**6.5.** Частица движется в статическом силовом поле по закону, показанному на рисунке. Найти зависимость силы и потенциала от координаты.

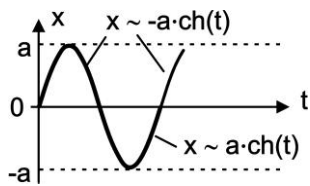

**6.6.** Найти закон движения частицы в поле $U = -\alpha x^4$ в случае, когда ее полная энергия равна нулю. Нарисовать траекторию частицы на фазовой плоскости.

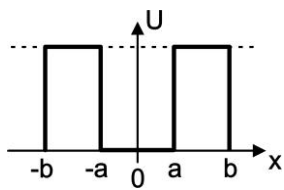

**6.7.** Нарисовать траектории частицы на фазовой плоскости при движении в одномерном поле с потенциалом, изображенным на рисунке.

**6.8.** Нарисовать траектории частицы на фазовой плоскости для следующих одномерных полей:

$$1)\ U(x) = \alpha^2\left(x^2 - b^2\right)^2, \quad 2)\ U(x) = -\alpha^2\left(x^2 - b^2\right)^2,$$

$$3)\ U(x) = U_0 \sin kx, \qquad 4)\ U(x) = \alpha^2\left(\frac{b^2}{x^2} - \frac{1}{x}\right).$$

**6.9.** Как изменится со временем форма и объем области в фазовом пространстве, занимаемой группой движущихся вдоль оси $x$ невзаимодействующих друг с другом частиц, помещенных в одномерный ящик с координатами стенок $x_1 = 0$, $x_2 = L$? Столкновения со стенками упругие. В начальный момент частицы занимали область $x_0$, $x_0 + \Delta x$ и $p_0$, $p_0 + \Delta p$.

**6.10.** Функция распределения частиц по скоростям и координатам $x$ в момент времени $t = 0$ имеет вид:

$$F(V_x, x, 0) = \begin{cases} \alpha\, n_0 \cdot \exp(-V_x^2 / V_0^2) & \text{при } x \le x_0 \\ 0 & \text{при } x > x_0 \end{cases}$$

Найти распределение плотности частиц по $x$ в момент времени $t$.

**6.11.** Как зависит период движения частицы в поле $U = \alpha|x|^\beta$ от ее энергии? $\alpha > 0$.

**6.12.** Желая определить распределение потенциала вдоль оси "черного ящика", экспериментатор пускает вдоль оси ионы с различными скоростями. Ионы, впущенные со скоростью $V$, возвращаются обратно через время $T = \alpha V^\beta$. Восстановите зависимость потенциала от координаты.

**6.13.** Изучая столкновения металлических шаров, студент обнару-жил, что время соприкосновения шаров $T = \alpha V^{-1/5}$, где $V$ – скорость шаров перед столкновением. Восстановите зависимость силы сопротивления шара деформации от величины деформации. Нарисуйте траектории сталкивающихся шаров на фазовой плоскости в системе центра масс.

**6.14.** Найдите зависимость периода колебаний частицы от энергии в одномерном поле с потенциалом $U(x) = \begin{cases} kx^2/2 & \text{при } |x| \le a, \\ \infty & \text{при } |x| > a. \end{cases}$

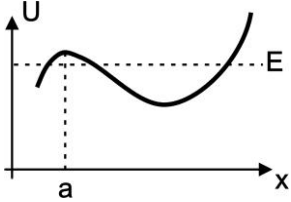

**6.15.** Определите, по какому закону обращается в бесконечность период движения частицы в поле, изображенном на рисунке, при приближении полной энергии частицы $E$ к $U(a)$, если производная потенциала $U'(a) = 0$, причем $U''(a) \ne 0$.

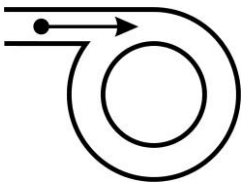

**6.16.** Через трубку, соединенную по касательной с тором, влетает шарик, испытывающий упругие отражения от стенок. Объяснить механизм возврата шарика в трубку, который должен иметь место согласно теореме Пуанкаре. Оценить время возврата в зависимости от размеров системы и начальных условий влета.

## 6.2. Движение с трением

**6.17.** Сила сопротивления, тормозящая моторную лодку, пропорциональна квадрату скорости. За какое время после выключения мотора ее скорость уменьшится вдвое? Какое расстояние при этом пройдет лодка?

**6.18.** На моторной лодке, плывшей против течения, выключили мотор. Нарисовать синхронные графики зависимости от времени скорости и смещения лодки относительно берега, если сопротивление воды пропорционально квадрату скорости лодки относительно воды. Изобразить движение на фазовой плоскости.

**6.19.** Найти закон движения тела при падении в поле тяжести. Сила сопротивления воздуха $F = -\alpha V^2$. Начальная скорость тела равна нулю.

**6.20.** На какую высоту и за какое время поднимется тело, брошенное вертикально вверх, если сопротивление воздуха пропорционально квадрату скорости?

**6.21.** Найти зависимость скорости тела от пройденного расстояния при падении в поле тяжести. Сила сопротивления воздуха пропор-циональна квадрату скорости. Начальная скорость тела равна нулю, максимальная скорость падения $V_{max}$.

**6.22.** Мячик подпрыгивает в поле тяжести над упругой плитой. Сила сопротивления воздуха пропорциональна квадрату скорости. Как меняются высота и время подскока мячика? Нарисовать синхронные графики скорости и координаты мячика в зависимости от времени, а также траекторию движения на фазовой плоскости.

**6.23.** Исследовать движение частицы в однородном силовом поле при линейной зависимости силы сопротивления от скорости. Нарисовать графики зависимости смещения, скорости и ускорения частицы от времени. Представить движение на фазовой плоскости.

**6.24.** Найти зависимость скорости движения кальмара от времени, если он затрачивает мощность $N$ и выбрасывает воду со скоростью $u$. Стартовая скорость кальмара равна нулю. Сила трения $F = -\alpha V$.

**6.25.** Тело массой $m$, подброшенное вертикально вверх с малой скоростью $V_1$, вернулось обратно со скоростью $V_2$. Сила сопротивления воздуха $F = -\alpha V$, ускорение свободного падения $g$. Сколько времени тело находилось в полете?

**6.26.** Тело массой $m$ брошено в поле тяжести с малой скоростью $V$ под углом $\theta$ к горизонту. Сила сопротивления $F = -\alpha V$. Какой будет скорость тела в верхней точке траектории?

**6.27.** На тонкое проволочное кольцо радиусом $R$ надета бусинка, которой сообщили скорость $V$. Найти путь, который пройдет бусинка до остановки, если коэффициент трения $\mu$. Кольцо неподвижно и расположено горизонтально в поле тяжести $g$.

**6.28.** Исследовать теоретически движение пробки в горлышке бутылки с подогретым шампанским. Нарисовать графики зависимости от времени скорости и смещения пробки.

**6.29.** Диск диаметром $D$, массой $M$ бомбардируется однородным потоком точечных пылинок массой $m \ll M$, плотностью $n$, скоростью $V$ (см. рисунок). За какое время первоначально неподвижный диск ускорится до скорости $u$? Удары пылинок упругие.

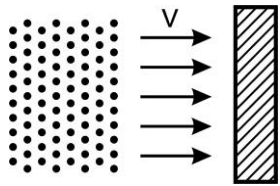

**6.30.** Диск диаметром $D$, массой $M$ летит через однородный поток (см. рисунок) движущихся навстречу точечных пылинок массой $m \ll M$.

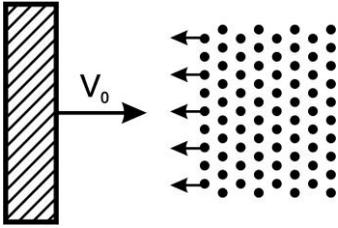

За какое время диск остановится? Плотность пылевого облака ρ, скорость *V*. Начальная скорость диска $V_0$, удары пылинок упругие.

**6.31.** Плоская тележка, двигавшаяся со скоростью *V,* попадает под вертикально падающий дождь (см. рисунок). Скорость капель *u*, средняя плотность дождя ρ, площадь горизонтальной поверхности тележки *S*. За какое время тележка остановится, если коэффициент трения колес о плоскость *μ*. Вода с тележки стекает, так что ее масса *M* остается постоянной**.**

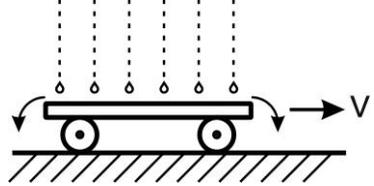

**6.32.** Тело движется в разреженном газе со скоростью, много большей тепловых скоростей молекул. Оценить силу трения для различных случаев взаимодействия молекулы с поверхностью тела. Что будет в релятивистском случае? Найти закон движения в случае, когда тело захватывает все столкнувшиеся с ним молекулы.

### 6.3. Движение с переменной массой

**6.33.** Однородный гибкий канат, висевший вертикально, падает на площадку весов. Нарисовать зависимость показаний весов от времени.

**6.34.** Однородная веревка соскальзывает со стола под действием силы тяжести через гладкую направляющую трубку (см. рисунок). Найти время соскальзывания, если в начальный момент длина свисающей части веревки равна четверти ее полной длины.

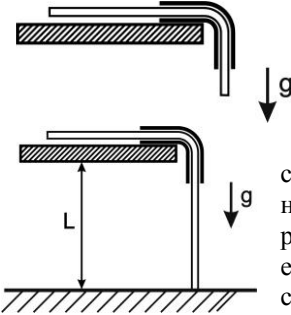

**6.35.** Однородная веревка длиной *2L* соскальзывает под действием силы тяжести через направляющую трубку со стола высотой *L* (см. рисунок). Найти максимальную скорость веревки, если в начальный момент она покоилась, а длина свисающей части была равна *L.*

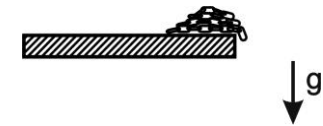

**6.36.** Свернутая в клубок тяжелая однородная цепь лежит на краю горизонтального стола, причем вначале одно звено цепи свешивается со стола. Под действием силы тяжести цепь начинает соскальзывать.

Принимая нулевые начальные условия, определить закон движения цепи. Считать, что звенья цепи поочередно приобретают только вертикальную скорость.

**6.37.** Свернутая в клубок тяжелая однородная цепь полной длиной $3H$ лежит на краю стола высотой $H$, так что ее один конец свешивается, касаясь пола (см. рисунок). Под действием силы тяжести цепь начинает соскальзывать. Найти зависимость скорости движущегося участка цепи от времени. За какое время цепь соскользнет со стола?

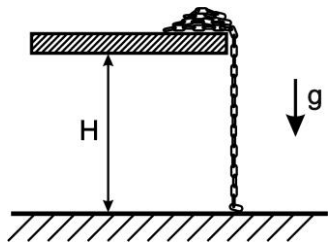

**6.38.** Какую массу газов должна ежесекундно выбрасывать ракета, чтобы оставаться неподвижной в поле тяжести?

**6.39.** При какой минимальной мощности двигателей ракета весом $10^3$ т сможет оторваться от стартового стола? Скорость истечения газов из сопла ракеты 2 км/с.

**6.40.** Ракета движется вверх с ускорением $a$ в однородном поле тяжести. За какое время масса ракеты уменьшится в два раза? Сопротивлением воздуха пренебречь. Скорость истечения газов $u$.

**6.41.** Определите путь, пройденный ракетой при ускорении из состояния покоя до скорости $u$, равной эффективной скорости истечения газов. Сопротивление воздуха и поле тяжести отсутствуют. Начальная масса ракеты $M_0$, секундный расход топлива $\mu$.

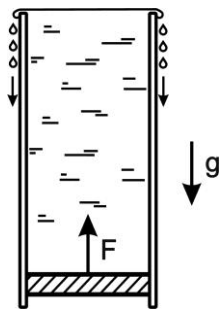

**6.42.** Вертикальная, открытая трубка длиной $L$ доверху заполнена водой массой $M$. Снизу трубка закрыта тонким поршнем массой $M$. Под действием постоянной силы $F$ поршень начинает двигаться вверх и вытесняет воду (см. рисунок). Найти скорость поршня в верхней точке. Поле тяжести $g$.

**6.43.** Найти закон движения сферической капли жидкости через неподвижный туман в поле тяжести. В начальный момент масса капли мала, а скорость равна нулю.

**6.44.** Найти закон движения капли жидкости под действием силы тяжести в пересыщенном паре. Скорость молекул пара много больше скорости капли. Скорость увеличения массы капли пропорциональна площади ее поверхности.

**6.45.** Ледяной метеорит сферической формы тормозится в атмосфере Земли. Сила трения о воздух пропорциональна его площади и скорости: $F = -\alpha SV$. Скорость испарения вещества метеорита пропорциональна его площади: $dM/dt = -\beta S$, причём в СО метеорита испарение изотропно. Найти зависимость скорости метеорита от времени, если его плотность $\rho$, а начальный радиус равен $R_0$. Силой тяжести пренебречь. Начальная скорость метеорита $V_0$.

**6.46.** На невесомый блок намотана тонкая верёвка массой $m$, длиной $L$. По верёвке начинает подниматься обезьянка массой $M$, при этом расстояние от неё до блока в процессе подъёма остаётся постоянным и равным $l$. Найти зависимость от времени скорости обезьянки относительно верёвки, если начальная длина свешивающейся части верёвки $l_0 > l$.

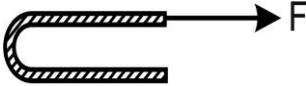

**6.47.** На гладком столе лежит нить, сложенная пополам. К одному из её концов приложена постоянная сила $F$ (см. рисунок). Описать движение нити.

**6.48.** Свернутая в клубок однородная цепь длины $L$, массы $m$ лежит на шероховатой поверхности с коэффициентом трения $\mu$. Цепь тянут за крайнее звено с постоянной горизонтальной силой $F > \mu\, mg$. Какой будет скорость цепи в момент, когда она полностью распрямится? Звенья цепи вовлекаются в движение поочерёдно.

**6.49.** Исследовать движение тележки, из которой вытекает вода через отверстие на дне (см. рисунок). Поверхность воды в тележке остаётся горизонтальной.

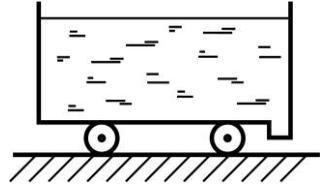

# 7. КОЛЕБАНИЯ

## 7.1. Свободные колебания

**7.1.** Определите частоту колебаний доски, положенной на два быстро вращающихся в противоположные стороны валика (см. рисунок), если расстояние между их осями $L$, коэффициент трения $\mu$.

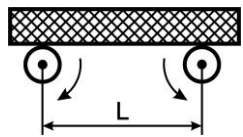

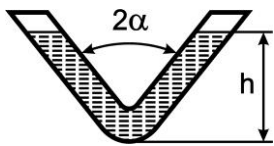

**7.2.** Найти частоту малых колебаний жидкости в трубке, показанной на рисунке. Высота трубки существенно больше радиуса закругления. Капиллярными эффектами пренебречь.

**7.3.** Через невесомый блок перекинута нерастяжимая нить массой $M$, длиной $L$, концы которой натянуты пружинами жесткостью $k$ (см. рисунок). Найти собственную частоту колебаний нити в поле тяжести. При каком условии колебания устойчивы? Трения нет.

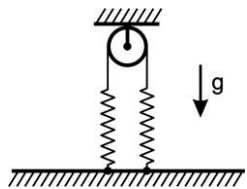

**7.4.** Определите время столкновения сильно накачанного мяча со стенкой. Масса мяча $m = 0{,}5$ кг, радиус $r = 0{,}1$ м. Избыточное давление $P = 7{,}5 \cdot 10^4$ Па в мяче в процессе удара меняется незначительно. Скорость мяча перпендикулярна стенке.

**7.5.** Какова амплитуда колебаний пружинных весов (см. рисунок) после быстрого падения каната длиной $L$, массой $m$? Начальная скорость каната равна нулю.

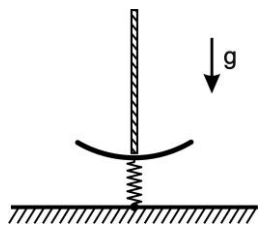

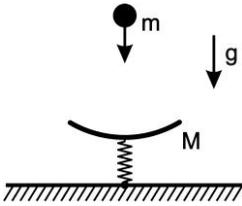

**7.6.** На неподвижную чашку весов массой *M* с пружиной жесткости *k* (см. рисунок) упал вертикально со скоростью *V* кусок пластилина массой *m*. Найти зависимость координаты чашки от времени после падения пластилина.

**7.7.** Точка массой *m*, несущая заряд *q*, может двигаться по вертикали в поле тяжести. Ниже на той же вертикали закреплен одноименный заряд *Q*. Найти частоту малых колебаний точки. В равновесии расстояние между точкой и зарядом Q равно *L*.

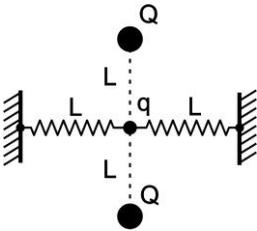

**7.8.** Найти частоту малых колебаний частицы с зарядом *q*, массой *m* вдоль линии, на которой закреплены заряды *Q* (см. рисунок). Пружины натянуты, их длина в нерастянутом состоянии $l < L$, коэффициент жёсткости *k*. Найти условие, при котором рассматриваемое положение равновесия станет неустойчивым.

**7.9.** В соленоиде с сильным однородным магнитным полем, равномерно заполненном покоившимися электронами, быстро включается электрическое поле, направленное вдоль оси. Потенциал поля зависит от координаты *x* вдоль оси по закону $U = ax^2$ (электроны ускоряются полем к центру $x = 0$). Найти зависимость плотности электронов от времени. Взаимодействием электронов между собой пренебречь. Как повлияет на результат конечный разброс начальных скоростей электронов?

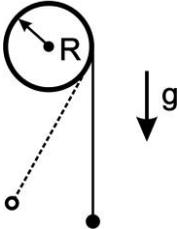

**7.10.** На закрепленный цилиндр радиусом *R* намотана нитка. Длина свисающей части нити *L*. На ней подвешен груз массой *m* (см. рисунок). Найти частоту малых колебаний.

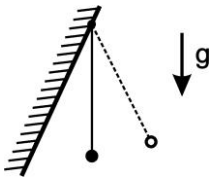

**7.11.** Грузик на нити подвешен к стенке с малым углом наклона к вертикали (см. рисунок). Определить зависимость периода колебаний грузика от угла отклонения от вертикали для случаев: а) удар о стенку упругий, б) удар неупругий.

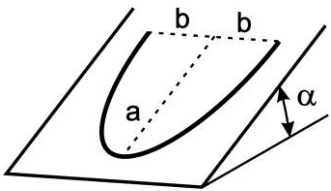

**7.12.** Частица может двигаться в поле тяжести по эллипсу, малая полуось которого **b** горизонтальна, а большая **a** составляет угол α с вертикалью (см. рисунок). Найти частоту малых колебаний частицы.

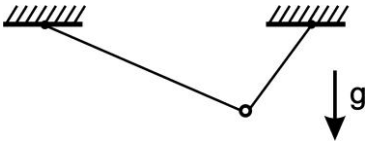

**7.13.** Бусинка надета на невесомую гладкую нить длиной *L*, концы которой закреплены на одинаковой высоте на расстоянии *d* друг от друга (см. рисунок). Найти частоту малых колебаний бусинки вдоль нити.

**7.14.** Найти период колебаний точки массой *m*, движущейся в поле тяжести в гладкой циклоидальной чашке $x = R(\varphi + \sin\varphi)$, $y = R(1 - \cos\varphi)$.

**7.15.** Гибкий однородный канат длины *L* скользит по горизонтали внутри узкой гладкой трубки со скоростью *u*. Сколько времени понадобится канату на преодоление V-образного углубления трубки (см. рисунок) с углом при вершине $2\alpha$ и глубиной $H = L\cos\alpha$?

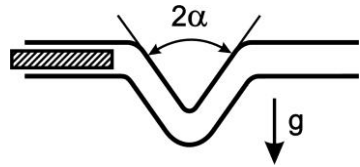

**7.16.** Гибкий однородный канат длины *L* скользит по горизонтали внутри узкой гладкой трубки со скоростью *u*. Сколько времени понадобится канату на преодоление Λ-образного подъема трубки (см. рисунок) с углом при вершине $2\alpha$ и высотой $H = L\cos\alpha$?

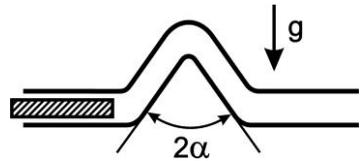

**7.17.** В точке максимального (или нулевого) отклонения математического маятника массой *M*, длиной *L* от него откололся кусочек массой *m*. Найти изменение энергии маятника, нарисовать график на фазовой плоскости.

## 7.2. Колебания с трением

**7.18.** Исследовать движение стрелки амперметра с нулем в центре шкалы после быстрого разрыва цепи постоянного тока, протекавшего через него. Учесть момент сил сухого трения в подшипниках оси. Нарисовать траекторию на фазовой плоскости для угла отклонения стрелки $\varphi$ и угловой скорости $\dot\varphi$.

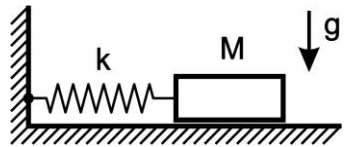

**7.19.** Лежащая на плоскости шайба массой *M* прикреплена к пружинке жесткости *k* (см. рисунок.) Шайбу сместили из положения равновесия на расстояние $A$ и отпустили. Какой путь пройдет шайба до остановки? Сила трения мала и пропорциональна скорости $F = -\alpha V$ $(\alpha^2 << k \cdot M)$.

**7.20.** При каком соотношении между сопротивлением $R$, индуктивностью $L$ и емкостью $C$ контура в цепи гальванометра осуществляется наиболее «оптимальный» апериодический режим демпфирования колебаний рамки гальванометра?

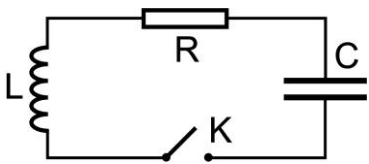

**7.21.** В контуре, изображенном на рисунке, конденсатор $C$ заряжен, причем $R^2 < 4L/C$. За какое время после замыкания ключа $K$ энергия, запасенная в контуре, уменьшится в 20 раз?

**7.22.** Собственная частота стрелки амперметра с нулем в центре шкалы 1 Гц, добротность при затухании колебаний $Q = 20$. За какое время амплитуда колебаний стрелки уменьшится в 20 раз, если выключить протекавший через амперметр постоянный ток?

### 7.3. Вынужденные колебания. Резонанс

**7.23.** Груз массой $m$ подвешен на пружине жесткостью $k$ и колеблется с амплитудой $A$ в поле тяжести. В момент, когда груз находится в крайнем нижнем положении, точку подвеса начинают двигать вверх с постоянной скоростью $V$. Найти зависимость координаты груза от времени.

**7.24.** Груз массой $m$ подвешен на пружине жесткости $k$. Найти амплитуду его колебаний, оставшихся после действия прямоугольного импульса силы амплитудой $F$, длительностью $\tau$. Сила направлена вдоль пружины. Начальная скорость груза была равна нулю.

**7.25.** Найти амплитуду отклонения указателя гальванометра магнитоэлектрической системы (в единицах показываемого тока), если через его рамку был пропущен прямоугольный импульс тока с амплитудой 1 А и длительностью $10^{-2}$ с. Период собственных колебаний рамки гальванометра 1 с. Затуханием колебаний пренебречь. Какой будет амплитуда колебаний при "треугольной" форме импульса тока?

**7.26.** Найти энергию, приобретенную осциллятором за все время действия силы $F = F_0 - F_0 \exp(-t/\tau)$. В начальный момент времени $t = 0$ энергия осциллятора была равна $E_0$ и он проходил через положение равновесия.

**7.27.** На осциллятор с трением, первоначально находившемся в равновесии, в течение времени $T$ действует сила $F$, при этом уравнение движения имеет вид: $\ddot{x} + 2\gamma\dot{x} + \omega^2 x = F/m$, $(0 \le t \le T)$. Найти энергию, перешедшую в тепло за все время движения осциллятора. При каком значении $T$ эта энергия достигает максимума? Нарисовать фазовые траектории осциллятора при различных соотношениях между $\omega$, $\gamma$ и $T$.

**7.28.** Осциллятор, состоящий из грузика массы $m$, подвешенного в поле тяжести на пружине, имеет период собственных колебаний $T$. Точку подвеса пружины двигают по вертикали по закону, показанному на рисунке. Какой будет энергия осциллятора через $N$ полупериодов, если вначале осциллятор покоился? Что будет при других начальных условиях?

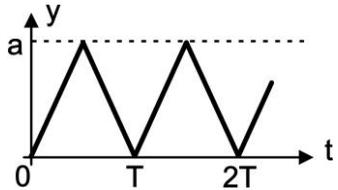

**7.29.** Определить амплитуду колебаний массы $m$, закрепленной на пружинке жесткостью $k$, оставшихся после воздействия внешней силы, график которой (один полупериод синусоиды) показан на рисунке. Период собственных колебаний осциллятора $T$ совпадает с периодом внешней силы. До включения силы осциллятор покоился. Каким будет результат, если внешняя сила действовала в течение $N$ полупериодов синусоиды?

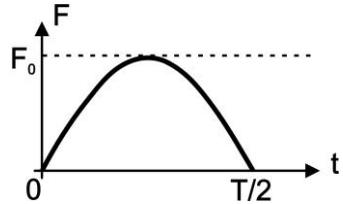

**7.30.** Период собственных колебаний рамки амперметра магнитоэлектрической системы равен 1 с, добротность $Q = 10$. Найти амплитуду установившихся колебаний стрелки (в единицах показываемого тока), если через амперметр пропускается синусоидальный ток с амплитудой 1 А и частотой 10 Гц. Какой будет амплитуда колебаний при резонансе?

**7.31.** Через амперметр магнитоэлектрической системы пропускают ток, изменяющийся по закону $I(t) = I_0 \sin^2 \Omega t$. Найти зависимость амплитуды установившихся колебаний от $\Omega$. Затухание считать малым. $I_0 = 1$ А. Амплитуду колебаний отсчитывать в амперах. Собственная частота $\omega_0$.

**7.32.** Через амперметр магнитоэлектрической системы пропускают выпрямленный синусоидальный ток с амплитудой 1 А, частотой 100 Гц. Какими будут среднее отклонение и амплитуда установившихся колебаний стрелки? Частота собственных колебаний стрелки 1 Гц.

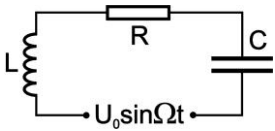

**7.33.** Найти амплитуду установившихся колебаний напряжения на конденсаторе и тока в контуре (см. рисунок), если на его вход подается переменное напряжение $U = U_0 \sin \Omega t$.

**7.34.** Шарик массой $m$ подвешен в поле тяжести на пружине жесткости $k$. Точка подвеса пружины движется по вертикали по закону $z = a \cos \Omega t$. Найти амплитуду установившихся малых колебаний шарика.

**7.35.** Точка подвеса математического маятника длиной $L$ движется по горизонтали по закону $x = b \cos \Omega t$. Найти угловую амплитуду установившихся малых колебаний.

**7.36.** Рессоры железнодорожного вагона прогибаются под его тяжестью на 4 см, расстояние между стыками железнодорожного полотна 25 м. При какой скорости поезда амплитуда вертикальных колебаний вагона будет максимальной?

**7.37.** Предположим, что радиус одного из колец Сатурна периодически изменяется со временем. Найти условия резонанса и проанализировать баланс энергии.

**7.38.** По какому закону нужно менять длину математического маятника, чтобы параметрическая раскачка колебаний была наиболее эффективной (качели)? Изобразить движение на фазовой плоскости.

## 7.4. Адиабатические инварианты

**7.39.** Упругий шарик подпрыгивает в поле тяжести над горизонтальной плитой. Поле тяжести медленно изменяется. Как меняется высота подскока шарика над плитой?

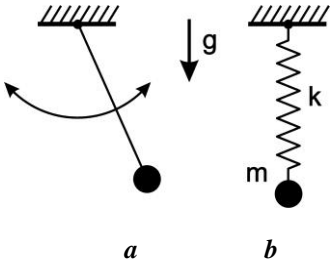

**7.40.** Масса осциллятора медленно возрастает. Как при этом изменяются амплитуда и период его колебаний? Рассмотреть оба случая, показанных на рисунке.

**7.41.** Масса осциллятора медленно уменьшается (например, из-за таяния). Как при этом изменяются амплитуда и период его колебаний? Рассмотреть оба случая, показанных на рисунке к предыдущей задаче.

**7.42.** Звезда теряет за счет излучения $10^{-9}$ часть своей массы в год. За какое время радиус круговой орбиты планеты, вращающейся вокруг звезды, изменится вдвое? Влиянием излучения на планету пренебречь.

**7.43.** Грузик, подвешенный на нити к наклонной стенке, совершает малые колебания относительно точки подвеса О в поле тяжести (см. рисунок). Стенка медленно поворачивается вокруг точки О и принимает вертикальное положение. Во сколько раз изменится угловая амплитуда колебаний грузика? Удар упругий, начальный угол наклона стенки больше амплитуды колебаний грузика.

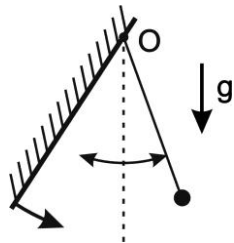

**7.44.** Вблизи точки подвеса математического маятника длиной $L$ в точке А забит гвоздь. Точку подвеса медленно поднимают по вертикали (см. рисунок). Во сколько раз изменится размах колебаний маятника к моменту, когда расстояние между гвоздем и точкой подвеса станет равно $L/2$ ?

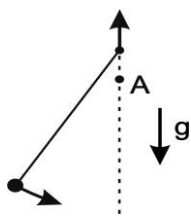

**7.45.** Частица движется со скоростью $V$ вдоль стороны $L$ в расположенном горизонтально прямоугольном ящике (см. рисунок), упруго отражаясь от его стенок. Ящик медленно поднимают за один конец, поворачивая вокруг ребра, перпендикулярного $L$. При каком угле наклона α дна ящика частица не будет достигать его верхней стенки?

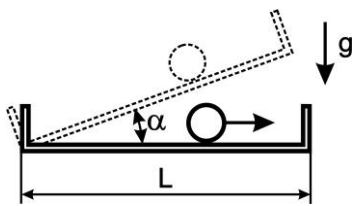

**7.46.** На нижнем конце невесомого стержня сидят два жука массой $m$ каждый. Стержень совершает малые колебания в поле тяжести (см. рисунок). Один из жуков медленно ползет вдоль стержня. Как изменится амплитуда колебаний к моменту, когда жук достигнет середины стержня?

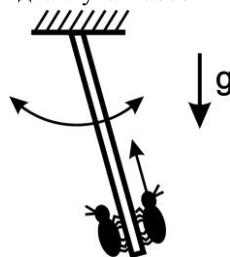

**7.47.** Найти период колебаний электронов вдоль оси $z$ при движении по винтовым траекториям в магнитной ловушке. Магнитное поле симметрично относительно оси $z$ и изменяется по закону $B_z = (1 + \lambda \, \text{th}^2 \, \alpha z)$. Вблизи оси компоненты поля $B_x = B_y = 0$. В центре ловушки скорость электронов $V_0$ составляет угол $\theta_0$ с осью $z$. Указание: воспользуйтесь адиабатической инвариантностью магнитного момента электрона $M = \dfrac{mV_\perp^2}{2B} = $ const, где $V_\perp$ − компонента скорости электрона, перпендикулярная «медленно» изменяющемуся в пространстве магнитному полю.

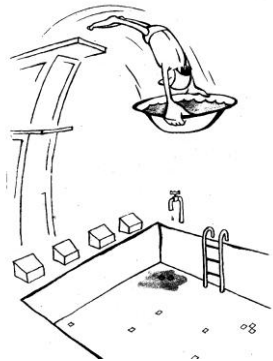

# 8. ДВИЖЕНИЕ В ЦЕНТРАЛЬНОМ ПОЛЕ

## 8.1. Потенциал поля

**8.1.** Нарисуйте график потенциала и напряженности поля тяготения Земли в зависимости от расстояния до центра Земли.

**8.2.** Найти давление в центре "жидкой планеты" шаровой формы. Плотность жидкости считать однородной и равной 5,5 г/см$^3$. Радиус планеты 6400 км. Как изменится результат, если не пренебрегать сжимаемостью жидкости при увеличении давления?

**8.3.** На какой высоте от поверхности планеты нужно включить тормозной двигатель космического аппарата, чтобы обеспечить мягкую посадку на поверхность? Спуск происходит по прямой, проходящей через центр планеты. Сила торможения $F$ постоянна. Сопротивлением воздуха и изменением массы аппарата пренебречь. Масса аппарата $m$, скорость вдали от Земли.

**8.4.** Внутри шара плотностью $\rho$ имеется сферическая полость, центр которой находится на расстоянии $\vec{a}$ от центра шара. Найти напряженность поля тяготения внутри полости.

**8.5.** Оценить относительное изменение ускорения свободного падения $\Delta g / g$ в шахте на глубине 10 км (где $g$ - ускорение на поверхности Земли). Средняя плотность вещества в земной коре в два раза меньше, чем средняя плотность Земли.

**8.6.** В результате сферически-симметричного взрыва однородного шара массой $M$, радиусом $R$ образуется множество мелких осколков. При какой минимальной суммарной кинетической энергии осколков они смогут разлететься на бесконечное расстояние от точки взрыва?

**8.7.** Сколько энергии выделится при гравитационном сжатии однородного шарового облака массой $M$, радиусом $R$ до радиуса $r$?

**8.8.** Оценить выход энергии при делении ядер урана $^{235}_{92}\text{U}$. Считать, что радиус ядра с числом нуклонов $A$ равен $R = R_0 \cdot A^{1/3}$, где $R_0 = 10^{-13}$ см.

**8.9.** Шарик массой $m$ находится в поле сил, имеющем точку равновесия $x = 0$, $y = 0$. Если его вывести из положения равновесия и отпустить, то он движется по закону $x = a\sin\omega t$, $y = b\cos\omega t\,t$. Найти зависимость силы $F = F\,(x,\,y)$ от координат.

**8.10.** На сферически-симметричную потенциальную "яму" радиусом $R$ и глубиной $U$ налетает плоский поток частиц с кинетической энергией $E$. В центре ямы расположена "липкая" сфера радиусом $a < R$. Найти зависимость сечения прилипания частиц к сфере от энергии частиц, построить график.

**8.11.** На сферически-симметричный потенциальный барьер радиусом $R$ и высотой $U$ налетает плоский поток частиц с кинетической энергией $E$. В центре барьера расположена "липкая" сфера радиусом $a < R$. Найти зависимость сечения прилипания частиц к сфере от энергии частиц, построить график.

**8.12.** По круговой орбите вокруг звезды массой $M$ движется планета массой $m \ll M$. В результате взрыва звезда сбрасывает массу $\alpha M$. Найти, при каком значении $\alpha$ планета покинет звезду. Считать, что сбрасываемая масса выходит за орбиту планеты сферически симметрично и мгновенно.

**8.13.** Вокруг звезды массой $M$ по круговой орбите радиусом $R$ двигался космический объект массой $m \ll M$. В результате взрыва объекта его осколки стали разлетаться изотропно с начальной скоростью $u$ (в системе отсчета объекта). Найти минимальное значение $u$, при котором не менее 25 % осколков покинет систему звезды.

**8.14.** У быстро вращающейся звезды массой $M$, радиусом $R$ взрывом сбрасывается тонкая верхняя шаровая оболочка. Какая часть сброшенного вещества вернется на звезду, если угловая скорость вращения звезды $\Omega$, а начальная радиальная скорость оболочки $u < \sqrt{2GM/R}$? Как выглядит расширяющаяся оболочка через большое время после взрыва? Масса оболочки много меньше массы звезды.

**8.15.** На спутник, движущийся по круговой орбите, действует слабая тормозящая сила $F = -\alpha V^2$. Найти зависимость скорости спутника от времени. За какое время радиус орбиты уменьшится на 2 %, если за месяц скорость спутника меняется на 1 %?

**8.16.** Оцените время жизни атома водорода с точки зрения классической физики, считая, что электрон вращается по круговой орбите радиусом $a_o = 5 \cdot 10^{-9}$ см и в единицу времени излучает энергию $\dfrac{2e^2 a^2}{3c^3}$, где $a$ – ускорение электрона, $e$ – его заряд, $c$ – скорость света (в системе CGSE).

**8.17.** Сферическая частичка радиусом 1 мм, массой $10^{-2}$ г движется по круговой орбите радиусом 500 св. с вокруг Солнца. Оцените силу торможения, обусловленную взаимодействием частички с излучением Солнца. Частичка разогревается и переизлучает тепло изотропно в своей системе отсчета. Мощность излучения Солнца $4 \cdot 10^{26}$ Вт.

**8.18.** Оценить силу торможения, обусловленную взаимодействием Земли с солнечным излучением. Расстояние Земля – Солнце $1{,}5 \cdot 10^{11}$ м, мощность излучения Солнца $4 \cdot 10^{26}$ Вт, радиус Земли $6{,}4 \cdot 10^6$ м. За какое время радиус орбиты Земли изменится в два раза?

## 8.2. Момент импульса. Центробежный потенциал

**8.19.** Через отверстие в гладком столе пропущена невесомая нить, к концам которой прикреплены массы $m_1$ и $m_2$. Масса $m_2$ лежит на расстоянии $r_0$ от отверстия (см. рисунок). Ей сообщают импульс $P$ перпендикулярно нити. Найти максимальное удаление массы $m_2$ от отверстия.

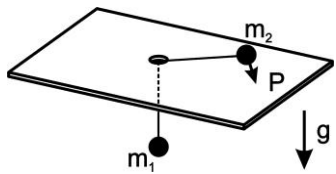

**8.20.** Электрон движется в плоскости, перпендикулярной положительно заряженной нити, по траектории, близкой к окружности радиуса $R$. Сила притяжения, действующая на электрон, равна $\alpha/r$, где $r$ – расстояние до нити. Найти период радиальных колебаний электрона.

**8.21.** Частица скользит без трения по стенке воронки (см. рисунок). В начальный момент частица находилась на высоте $h$ и двигалась горизонтально со скоростью $V$. При какой минимальной скорости $V$ частица не провалится в воронку, отверстие которой имеет радиус $\rho_0$?

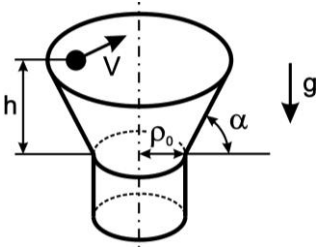

К задаче 8.21.

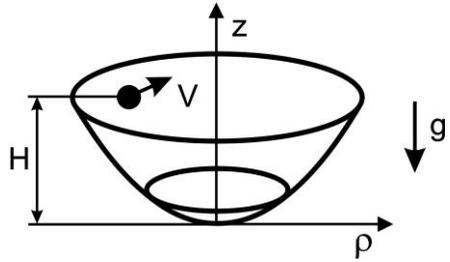

К задаче 8.22.

**8.22.** Частица движется без трения по поверхности параболической чашки, описываемой в цилиндрической системе координат уравнением $z = \alpha \rho^2$. Поле тяжести направлено вдоль оси $z$. На высоте $H$ скорость частицы $V$ была горизонтальна (см. рисунок). Найти границы движения частицы.

**8.23.** Найти период движения частицы массой $m$ в центральном поле с потенциалом $U = \alpha r^2$ ($\alpha > 0$).

**8.24.** Найти сечение падения потока метеоритов на Землю. Скорость метеоритов вдали от Земли $V_\infty$.

**8.25.** На большом расстоянии $R$ от Земли взорвался космический объект массой $M$. Осколки разлетелись сферически симметрично со скоростью $V$. Какая масса продуктов взрыва попадет на Землю? Радиус Земли $R_3$, вторая космическая скорость $V_2$.

**8.26.** Найти сечение падения частиц энергией $E$ на сферу радиусом $R$, находящуюся в центре поля с потенциалом отталкивания $U = \alpha/r$.

**8.27.** Найти сечение падения в центр поля притяжения $U = -\alpha/r^4$.

**8.28.** Найти сечение падения на сферу радиусом $R$, находящуюся в центре поля притяжения с потенциалом $U = -\alpha \cdot r^{-3/2}$.

**8.29.** Найти сечение падения на сферу радиусом $R$, находящуюся в центре поля с потенциалом $U = -\alpha/r - \beta/r^2$. $\alpha, \beta > 0$. Скорость частиц на бесконечности $V_\infty > \alpha(m\beta)^{-1/2}$.

**8.30.** Найти сечение падения в центр поля $U = -\alpha/r^n + \beta/r^2$. $\alpha, \beta > 0$.

**8.31.** Точка массой $m$ движется в центральном поле, причем ее скорость $V = \alpha/r$, где $\alpha = $ const. Найти зависимость силы $F$ от расстояния до центра поля $r$ и траекторию точки.

**8.32.** Частице массы М, находящейся в центральном поле $U = \alpha\, r^{-2}$ на расстоянии $r_0$ от центра, сообщили скорость $\vec{V}_0 \perp \vec{r}_0$. Найти уравнение траектории частицы.

**8.33.** При движении в центральном поле скорость частицы массой $m$ изменяется по закону $V = \alpha\, r^{-1/2}$. Восстановите зависимость силы от расстояния до центра поля $r$. Найдите уравнение траектории частицы в случае, когда ее максимальное приближение к центру поля имеет величину $r_0$.

**8.34.** Частица движется в центральном поле по дуге окружности радиуса $R$, проходящей через центр поля. Доказать, что потенциал поля имеет вид $U = -\alpha r^{-4}\ (\alpha > 0)$. Найти начальные условия такого движения.

**8.35.** Найти время падения массы $m$ в центр поля $U = -\alpha/r^6$ с расстояния $R$, если ее полная энергия равна нулю, а начальная скорость перпендикулярна направлению на центр.

**8.36.** Найти время падения массы $m$ в центр поля $U = -\alpha/r^4$ с расстояния $R$, если ее полная энергия равна нулю, а начальная скорость перпендикулярна направлению на центр.

**8.37.** Найти период малых радиальных колебаний релятивистской частицы вблизи круговой орбиты при движении в поле с потенциалом
$$U = -\alpha/r - \beta/r^2.$$

## 8.3.  Кулоновское поле. Законы Кеплера

**8.38.** Какой должна быть минимальная скорость ракеты при выходе из атмосферы Земли, чтобы она смогла покинуть Солнечную систему без дополнительного ускорения?

**8.39.** Космический корабль должен покинуть Солнечную систему в определенном направлении. Какова минимальная скорость корабля при выходе из атмосферы Земли, необходимая для этого?

**8.40.** Космический корабль приближается к Луне по параболической траектории, почти касающейся поверхности Луны. Чтобы перейти на круговую орбиту, в момент наибольшего сближения включают тормозной ионный двигатель, выбрасывающий поток ионов цезия $^{133}\text{Cs}^+$ (ускоряющее напряжение 1 кВ). Какую часть общей массы должен потерять корабль? Радиус Луны 1740 км, ускорение силы тяжести $g\,/6$.

**8.41.** Находящийся на круговой орбите космический корабль тангенциальной добавкой скорости переводят на гиперболическую орбиту со скоростью на бесконечности. При каком радиусе начальной круговой орбиты эта добавка скорости минимальна?

**8.42.** Оцените, с какой минимальной скоростью нужно стартовать с поверхности Луны, чтобы вернуться на Землю? Ускорение свободного падения на Луне $g/6$, скорость движения Луны по орбите 1 км/с. Радиус Луны 1740 км.

**8.43.** Баллистическая ракета стартует с Земли и продолжает свободный полет по траектории, апогей которой равен радиусу орбиты Луны. Какую максимальную скорость относительно Солнца сможет приобрести ракета при "правильном" использовании поля тяготения Луны ("гравитационный" маневр)?

**8.44.** Оценить поправку ко второй космической скорости, связанную с наличием Луны.

**8.45.** Какой должна быть минимальная скорость запуска тела с поверхности Луны, чтобы оно улетело за пределы Солнечной системы? Орбитальная скорость Земли 30 км/с, Луны – 1 км/с, ускорение свободного падения на поверхности Луны в шесть раз меньше, чем на Земле, радиус Луны 1740 км.

**8.46.** Комета Галлея движется по сильно вытянутой орбите с минимальным расстоянием до Солнца 0,6 а. е. Во сколько раз максимальная скорость кометы больше скорости движения Земли вокруг Солнца?

**8.47.** Оценить скорость движения предметов внутри орбитальной станции, двигающейся по околоземной орбите.

**8.48.** В перигее величиной $r_{мин}$ скорость спутника $V$. При каком касательном приросте скорости в перигее высота апогея увеличится на 1 %?

**8.49.** Орбитальная станция движется по круговой траектории на расстоянии 200 км от поверхности Земли. Какую наименьшую дополнительную скорость надо сообщить станции, чтобы ее максимальное удаление от Земли достигло 210 км?

**8.50.** С какой минимальной скоростью должен покинуть атмосферу Земли космический корабль, направляющийся к Марсу и стартующий по касательной к орбите Земли? Каким будет расстояние от Земли до Марса при посадке корабля на Марс? Радиус орбиты Марса 1,52 а. е. Какова минимальная начальная скорость при полете на Венеру? Радиус орбиты Венеры 0,72 а. е.

**8.51.** Спутник движется по околоземной круговой орбите радиусом $r$. Какую радиальную добавку скорости ему нужно сообщить, чтобы его орбита стала эллиптической с перигеем $r_1$?

**8.52.** Баллистическую ракету запускают с Северного полюса, так что после выхода из атмосферы и выключения двигателей она имеет скорость $V_0$ и угол вылета $\theta$ по отношению к горизонту. При каком соотношении между $V_0$ и $\theta$ ракета достигнет Южного полюса?

**8.53.** Как следует запускать баллистическую ракету на экваторе, чтобы она попала на Северный полюс? Найти связь между величиной начальной скорости и направлением запуска.

**8.54.** Требуется вывести космический корабль на околосолнечную орбиту с перигелием 0,001 а.е. и периодом обращения один год. С какой скоростью и в каком направлении относительно линии Земля – Солнце нужно запустить такой корабль с Земли?

**8.55.** Астероид, вращавшийся вокруг Солнца по круговой орбите со скоростью 20 км/с, за счет столкновения с метеоритом получил добавку тангенциальной скорости 20 км/с. С какой скоростью и под каким углом к первоначальной скорости астероид покинет пределы Солнечной системы?

**8.56.** Астероид движется вокруг Солнца по эллиптической орбите с апогелием 2,8 а. е. и перигелием 1,01 а. е. При каком минимальном относительном изменении скорости в апогелии астероид столкнется с Землей? Какова при этом максимальная относительная скорость Земли и астероида при встрече? Орбита астероида лежит в плоскости орбиты Земли.

**8.57.** С какой скоростью спутник должен покинуть атмосферу Земли, чтобы выйти на орбиту вокруг Солнца с перигелием $r_1 = 0,2$ а. е. и апогелием $r_2 = 1,8$ а. е.? Плоскость орбиты спутника лежит в плоскости орбиты Земли.

**8.58.** На космическом аппарате, движущемся по круговой орбите радиусом 1 а. е. вокруг Солнца, ставится идеально отражающий излучение парус, ориентированный перпендикулярно лучам Солнца. Найти минимальную площадь паруса, необходимую для того, чтобы покинуть солнечную систему. Масса аппарата $m = 10$ т, масса Солнца $M = 2 \cdot 10^{30}$ кг, полная мощность излучения $N = 3,86 \cdot 10^{26}$ Вт. Какова минимальная площадь паруса для полета к орбите Марса (радиус орбиты 1,52 а.е.)?

**8.59.** Одно тело движется по параболе, другое – по окружности. В результате неупругого столкновения в перигее они слипаются. Найти траекторию образовавшегося тела. Массы тел одинаковы.

**8.60.** Три звезды одинаковой массы $M$, находящиеся в вершинах равностороннего треугольника со стороной $d$, движутся вдоль его сторон с одинаковыми начальными скоростями $V = \sqrt{GM/d}$ (см. рис. *a*). Каким будет минимальное расстояние между звездами в процессе движения? Каким будет максимальное расстояние между звездами, если скорости звезд были направлены перпендикулярно сторонам треугольника (см. рис. *b*)?

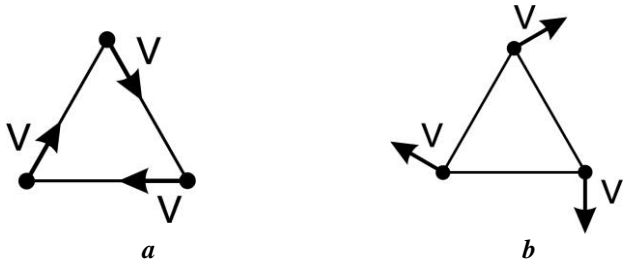

*a*　　　　　　　*b*

**8.61.** Сколько лет нужно ожидать возвращения кометы, удаляющейся от Солнца на 35 а. е.? Перигелий кометы 0,6 а. е.

**8.62.** При каком изменении скорости движения Земли продолжительность года увеличится в два раза?

**8.63.** Спутник движется по окружности радиусом $R$ с периодом $T$. За короткое время скорость спутника была увеличена в $k$ раз без изменения ее направления. Найти максимальное удаление спутника от центра Земли и новый период обращения.

**8.64.** Спутник, двигавшийся по круговой орбите, получил радиальную добавку скорости $\Delta V$. Как изменился период обращения спутника? Что будет, если добавка скорости перпендикулярна плоскости орбиты?

**8.65.** Два спутника движутся друг за другом на расстоянии 45 км по общей круговой орбите вблизи Земли. Чтобы состыковаться, спутники должны сблизиться и двигаться по общей орбите. Сколько раз нужно включить двигатель отстающего спутника, чтобы осуществить этот маневр наиболее экономично? Как зависит время сближения спутников от величины добавки к скорости? Двигатель сообщает спутнику импульс, перпендикулярный радиусу орбиты, а его каждое включение изменяет скорость спутника не более, чем на 8 км/ч.

**8.66.** За какое время Земля упадет на Солнце, если остановить ее движение по орбите?

**8.67.** Оценить время, через которое возвратится баллистическая ракета, запущенная с поверхности Земли со скоростью 10 км/с. Сопротивлением атмосферы пренебречь.

**8.68**. С какой начальной скоростью добрый молодец подбросил дубинку, если она вернулась на Землю через трое суток?

**8.69.** Как изменится период обращения Земли вокруг Солнца после неупругого столкновения с осколком, масса которого в $10^6$ раз меньше массы Земли? Относительно Солнца осколок двигался по параболе и перед столкновением летел под углом $\alpha$ к скорости Земли.

**8.70.** Найти закон движения частицы по параболической траектории в поле с потенциалом $U = -\alpha/r$.

**8.71.** Найти траекторию частицы в поле

$$U = \begin{cases} -\alpha/r & \text{при} \quad r \ge R \\ -\dfrac{3\alpha}{2r} + \dfrac{\alpha r^2}{2R^3} & \text{при} \quad r < R. \end{cases}$$

**8.72.** Определить траекторию движения релятивистского электрона в поле закрепленного ядра с зарядом $Ze$. Исследовать траектории для случая $L < Ze^2/c$ и $L \ge Ze^2/c$, где $L$ – момент импульса электрона. Найти скорость прецессии орбиты, обусловленной релятивистскими поправками.

### 8.4. Задача двух тел

**8.73.** Найти период малых продольных колебаний осциллятора, состоящего из двух масс $m$ и $M$, закрепленных на концах пружины жесткостью $k$.

**8.74.** Как изменится скорость хода часов с крутильным маятником, если их снять с пульта космического корабля и оставить свободно парить в кабине? От каких параметров и как будет зависеть это изменение?

**8.75.** Через невесомый блок перекинута нерастяжимая нить, к концам которой через пружины жесткостью $k$ подвешены грузики массой $m$ и $M$. Найти частоту малых колебаний грузиков в поле тяжести. Трения нет.

**8.76.** Два шарика массами $m$ и $M$ соединены пружиной жесткости $k$. Шарики заряжают одноименными зарядами, так что пружина растягивается в $\alpha$ раз (пружина электрическое поле не возмущает). Найти частоту малых продольных колебаний системы.

**8.77.** В линейной цепочке из трех масс $m$, $2m$ и $m$, соединенных пружинами жесткостью $k$, возбуждены симметричные колебания ампли-тудой $A$ (см. рисунок). В фазе сжатия пружин средняя масса разваливается на две равные части. Найти частоту и амплитуду колебаний новых цепочек.

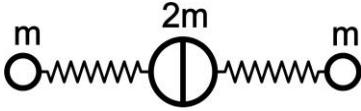

**8.78.** Частица массой $M$, скоростью $V$ испытывает лобовое упругое столкновение с первоначально неподвижной частицей массой $2M$, соединенной невесомой пружиной жесткостью $k$ с другой массой $M$ (см.

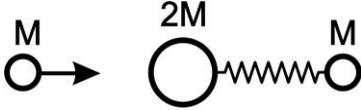

рисунок). Найти законы движения частиц после столкновения. В момент удара пружина была нерастянута и имела длину $L$.

**8.79.** Кусок пластилина массой $m$, скоростью $V$ испытывает лобовое неупругое столкновение с первоначально неподвижной частицей такой же массы, соединенной с другой массой $2m$ нерастянутой невесомой пружиной длиной $L$, жесткостью $k$ (см. рисунок). Найти законы движения частиц. Силу сцепления пластилина с частицей $m$ считать равной нулю.

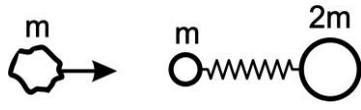

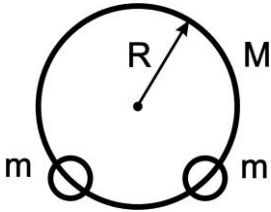

**8.80.** Две одноименно заряженные бусинки массой $m$ и зарядом $q$ каждая надеты на жесткое кольцо массой $M$, которое лежит на гладком столе (см. рисунок). Каким будет период малых колебаний системы, если бусинки сместить от положения равновесия и отпустить? Радиус кольца $R$.

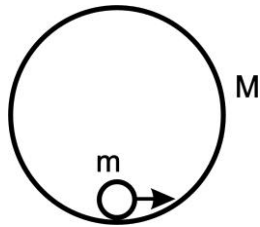

**8.81.** Внутри первоначально неподвижной гладкой сферы радиусом $R$, массой $M$ лежит шарик радиусом $r$, массой $m$ (см. рисунок). Шарику сообщается начальная поступательная скорость $V$, так что он начинает скользить по внутренней поверхности сферы. Описать дальнейшее движение системы. Силы трения и тяготения не учитывать.

**8.82.** Найти энергию связи атома позитрония, состоящего из электрона и позитрона, движущихся по круговой орбите радиусом $R$ вокруг общего центра масс. Насколько эта энергия отличается от энергии связи атома водорода в случае, когда радиус орбиты электрона в позитронии и в атоме водорода одинаковы? Заряд электрона $e$.

**8.83.** Через какое время столкнутся две точки с разными массами, начавшие двигаться из состояния покоя под действием силы взаимного гравитационного притяжения?

**8.84.** Частица массой $m$, скоростью $V$ налетает на первоначально покоящуюся частицу массой $M$. Прицельный параметр столкновения $\rho$. Найти минимальное расстояние между частицами, если потенциал взаимодействия: а) $U = \alpha^2/r^2$; б) $U = \alpha^2/r^4$.

**8.85.** Компоненты двойной звезды имеют массы $M$ и $2M$. Скорости звезд в начальный момент направлены перпендикулярно отрезку $d$, соединяющему их центры, и равны $2V$ и $V$ соответственно. Нарисовать возможные траектории звезд. Сформулировать условие финитности движения для этого случая. Вычислить период их движения, а также максимальное и минимальное расстояния между ними.

**8.86.** Найти полную массу системы "двойная звезда" по периоду обращения, минимальному и максимальному расстояниями между составляющими ее звездами.

**8.87.** Найти расстояние между компонентами двойной звезды, если их суммарная масса равна удвоенной массе Солнца, а звезды вращаются по круговым орбитам вокруг общего центра масс с периодом два года. Расстояние от Земли до Солнца $1{,}5 \cdot 10^8$ км.

**8.88.** Масса шарового астероида $M$, радиус $R$. Какой должна быть минимальная скорость у шара массой $m \le M$, радиусом $r < R$, запускаемого с поверхности астероида, чтобы он не вернулся на астероид?

**8.89.** Две звезды массами $m_1$ и $m_2$ двигаются по окружностям вокруг общего центра масс. У звезды массой $m_2$ в результате сферически-симметричного взрыва сбрасывается внешняя оболочка массой $qm_2$, которая, расширяясь с большой скоростью, быстро уходит за пределы двойной системы. При каком значении $q$ двойная система перестанет быть связанной гравитационными силами?

**8.90.** Две звезды с массами $m$ и $2m$ движутся по окружностям вокруг общего центра масс на расстоянии $d$ друг от друга. У звезды с массой $2m$ сферически-симметричным взрывом сбрасывается половина массы. Сброшенная оболочка, быстро расширяясь, покидает двойную систему. Каким будет новое максимальное расстояние между звездами? Во сколько раз изменится период их обращения?

## 8.5. Рассеяние частиц

**8.91.** Найти зависимость угла рассеяния точечных частиц на абсолютно упругой сфере радиусом $R$ от прицельного параметра $\rho$.

**8.92.** Найти сечение рассеяния на угол, больший $90^0$, при упругом столкновении точечной частицы массой $m$ с первоначально неподвижной сферой радиусом $R$, массой $2m$.

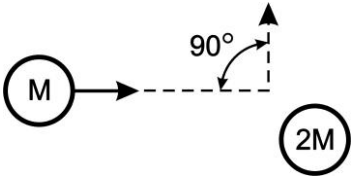

**8.93.** Сфера радиусом $R$, массой $M$ при упругом столкновении с первоначально неподвижным шаром радиусом $R$ и массой $2M$ рассеивается на угол $90^0$ (см. рисунок). Найти прицельный параметр этого столкновения.

**8.94.** Точечная частица массой $m$ и энергией $T$ упруго рассеивается на первоначально неподвижном шаре массой $M$. Какая часть энергии в среднем передается шару за одно столкновение?

**8.95.** Найти зависимость угла рассеяния частиц с энергией $E$ от прицельного параметра $\rho$ при рассеянии на сферически-симметричном потенциальном барьере высотой $U$ и радиусом $R$. Нарисовать график зависимости максимального угла рассеяния от энергии частиц при фиксированной высоте барьера. Нарисовать зависимость максимального угла рассеяния от высоты барьера при фиксированной энергии частиц.

**8.96.** Найти зависимость угла рассеяния от прицельного параметра при рассеянии частиц с кинетической энергией $E$ на сферической потенциальной "яме" глубиной $U$. Нарисовать график зависимости максимального угла рассеяния от энергии частиц при фиксированной глубине ямы. Нарисовать зависимость максимального угла рассеяния от глубины ямы при фиксированной энергии частиц.

**8.97.** Найти зависимость угла рассеяния от прицельного параметра для быстрых электронов, пролетающих мимо тонкой заряженной проволочки перпендикулярно ее оси (напряженность электрического поля проволочки обратно пропорциональна расстоянию от нее).

**8.98.** Найти сечение рассеяния на угол, больший $90^0$, при столкновении электрона с энергией $T = 10$ кэВ с неподвижным протоном. Как изменится результат, если протон не закреплен?

**8.99.** Найти сечение рассеяния на угол, больший $90^0$, при упругом столкновении протона с энергией $T = 10$ эВ с летящим навстречу протоном такой же энергии.

**8.100.** Плоский поток частиц рассеивается на отталкивающем кулоновском потенциале. Найти область, в которую частицы попасть не могут.

**8.101.** Найти зависимость энергии, переданной покоившемуся протону нерелятивистским электроном от прицельного параметра. Каким будет результат при столкновении ядер дейтерия и гелия? Столкновения упругие.

**8.102.** Найти сечение упругого рассеяния электрона с кинетической энергией 1 МэВ на угол, больший $10^{-2}$ радиан, при пролете мимо первоначально покоившегося протона. Определить максимальную и минимальную энергию, переданную протону при рассеянии.

**8.103.** Оценить сечение "ионизации" (отрыва планеты) Солнечной системы быстрой звездой. Скорость звезды $u$ много больше орбитальной скорости планеты $V_0$.

**8.104.** Мюон с кинетической энергией $T$ упруго рассеивается на первоначально покоившемся протоне. Прицельный параметр столкновения $\rho = 10^{-8}$ см. Найти энергии и направления разлета частиц после столкновения, если: а) $T = 10$ эВ; б) $T = 10$ МэВ. Определить минимальное расстояние между частицами в обоих случаях.

**8.105.** Найти зависимость угла рассеяния от прицельного параметра в поле $U = \alpha / r + \beta / r^2$, где $\alpha, \beta > 0$.

**8.106.** Найти уравнение траектории частицы массой $m$, движущейся в поле $U = \alpha r^2 + \beta / r^2$, где $\alpha, \beta > 0$.

**8.107.** Описать качественно характер движения и вид траектории частицы в поле $U = -\dfrac{\alpha}{r^{1/2}} - \dfrac{\beta}{r^{7/2}}$.

**8.108.** Описать качественно характер движения частицы в поле $U = \alpha r^{-1} \exp\left(-r/r_0\right)$ при различных значениях момента импульса и энергии частицы. $\alpha > 0$.

**8.109.** Два встречных цилиндрических сгустка нейтральных частиц имеют радиус $a$, длину $L$, скорость $V$. В каждом сгустке $N$ частиц, радиус которых $r$ много меньше среднего расстояния между частицами в сгустке. Найти полное число столкновений частиц за время прохождения сгустков сквозь друг друга.

**8.110.** Пучок α-частиц с энергией 10 МэВ проходит через золотую фольгу толщиной 10 мк. За час происходит в среднем одно рассеяние на угол, больший $90^0$. Найти интенсивность пучка α-частиц.

**8.111.** Оценить скорость потери энергии легким упругим шариком массой $m$ на покоящихся тяжелых шариках с таким же радиусом и массой $M \gg m$.

**8.112.** Найти толщину графитового замедлителя, понижающего среднюю кинетическую энергию нейтронов с 5 МэВ до 0,5 кэВ. Считать, что потери энергии происходят за счет упругих столкновений с ядрами углерода $C^{12}$ радиусом $3,5 \cdot 10^{-13}$ см, рассматриваемыми как покоящиеся упругие шарики.

**8.113.** Пучок быстрых отрицательных ионов проходит через перезарядную мишень с интегральной плотностью молекул $\int n\,dx = N$ мол/ см$^2$. Сечение перезарядки отрицательных ионов в атомы $\sigma_{-0}$ см$^2$/мол, сечение перезарядки атомов в протоны $\sigma_{0+}$ см$^2$/мол. Какая доля пучка отрицательных ионов выйдет из мишени в виде отрицательных ионов, атомов и протонов? При какой толщине мишени выход атомов максимален?

# 9. ДВИЖЕНИЕ ТВЕРДОГО ТЕЛА

## 9.1. Равновесие тел

**9.1**. Найти натяжение кольцевой цепочки, надетой на гладкий конус с углом при вершине $\alpha$.

**9.2.** Найти натяжение кольцевой цепочки радиусом $r$, весом $P$, надетой на гладкую сферу радиусом $R$ (см. рисунок).

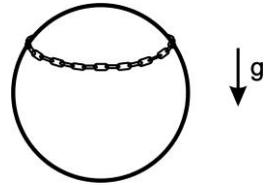

**9.3.** Обруч радиусом $R$, весом $P$ лежит горизонтально в параболической чашке, описываемой уравнением $y = \alpha \rho^2$. Найти силу упругого сжатия обруча $T$.

**9.4.** Найти силу упругого сжатия обруча радиусом $r$, весом $P$, лежащего горизонтально внутри гладкой сферической чашки радиуса $R$.

**9.5.** Найти силу, сжимающую невесомый стержень BD в системе, показанной на рисунке. Длина каждого стержня равна $L$, вес $P$. Стержни соединены шарнирами и образуют квадрат.

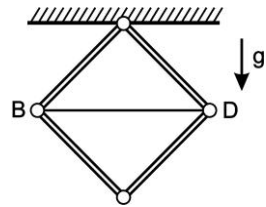

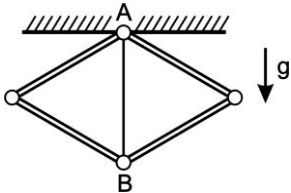

**9.6.** Ромб, составленный из четырех шарнирно закрепленных стержней весом $P$ каждый, подвешен за вершину (см. рисунок). Найти натяжение невесомой нити, соединяющей верхнюю и нижнюю точки AB ромба.

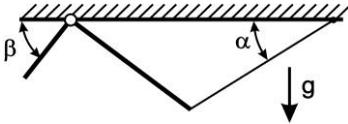

**9.7.** Однородная пластина длиной $3L$ и весом $3P$ согнута под прямым углом и подвешена, как показано на рисунке. Найти натяжение невесомой нити. При каком угле β нить нужно заменить невесомым стержнем?

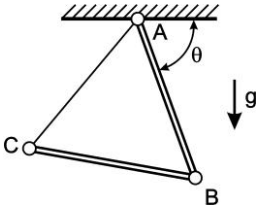

**9.8.** Одинаковые стержни AB и BC длиной $L$, весом $P$ соединены шарнирно. Точки A и C соединены невесомой нитью длины $L$. Найти натяжение нити и угол θ, образуемый стержнем AB с горизонталью в положении равновесия.

**9.9.** Однородный стержень длиной $L$, весом $P$ может скользить своими концами без трения по параболе $y = \alpha x^2$. Найти положения равновесия стержня и исследовать их устойчивость.

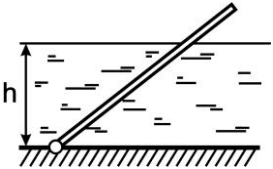

**9.10.** Нижний конец тонкой деревянной палочки длиной $L$ шарнирно закреплен на дне бассейна (см. рисунок). Глубина воды $h < L$. Найти положения равновесия и исследовать их устойчивость.

**9.11.** Тонкая деревянная палочка длиной $L$ шарнирно подвешена за один конец над поверхностью воды (см. рисунок, $h < L$). Найти положения равновесия палочки и исследовать их устойчивость.

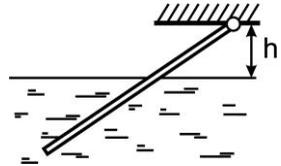

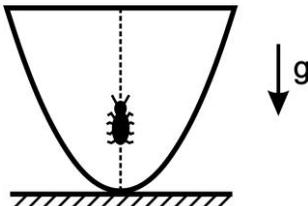

**9.12.** Невесомая плоская параболическая качалка высотой $H = 20$ см и шириной $L = 40$ см установлена вертикально на горизонтальной поверхности в поле тяжести и может качаться в своей плоскости (см. рисунок). По вертикальной оси качалки снизу вверх ползет маленький жук. До какой высоты он должен доползти, чтобы равновесие качалки стало неустойчивым и она могла наклониться?

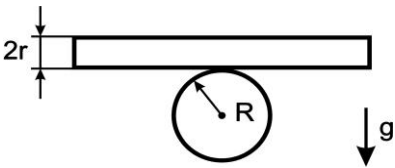

**9.13.** Карандаш радиусом $r$ удерживается горизонтально в равновесии на стержне радиусом $R$ в поле тяжести (см. рисунок). Оси карандаша и стержня перпендикулярны, коэффициент трения скольжения μ. При каком максимальном угле отклонения α карандаша от горизонтали он еще вернется в положение равновесия?

## 9.2. Вращение с неизменной ориентацией оси.
## Момент инерции, момент импульса

**9.14.** Крест состоит из однородных стержней, скрепленных посредине под углом α**.** Найдите его момент инерции относительно конца одного из стержней. Ось вращения перпендикулярна плоскости креста.

**9.15.** Из однородной квадратной пластины стороной *a* вырезали квадрат стороной *a/2* (см. рисунок). Масса полученной фигуры *m*. Найти момент инерции фигуры относительно оси, перпендикулярной плоскости рисунка и проходящей через её центр тяжести.

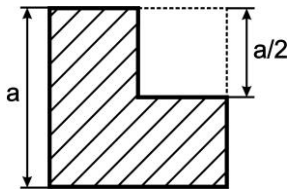

**9.16.** При каком соотношении между радиусом и высотой конуса его центральные главные моменты инерции будут одинаковыми?

**9.17.** Стержень вращается с угловой скоростью ω, причем ось вращения проходит через середину стержня, образуя с ним угол *α*. Найти величину и направление момента импульса, а также кинетическую энергию стержня.

**9.18.** На поверхность Земли выпадает метеорная пыль. Поток ее изотропен, плотность потока μ. Найти зависимость продолжительности суток от времени.

**9.19.** Оценить период вращения Солнца, если бы оно превратилось в нейтронную звезду с плотностью $10^{14}$ г/см$^3$. Средняя плотность Солнца 1,4 г/см$^3$, период вращения $2 \cdot 10^6$ с.

**9.20.** Масса вращающейся звезды уменьшается за счет быстрого истечения вещества в пространство. Как изменяется угловая скорость вращения звезды при уменьшении ее радиуса? Звезду считать однородным твердым шаром. Разлет вещества сферически симметричен в системе отсчета звезды.

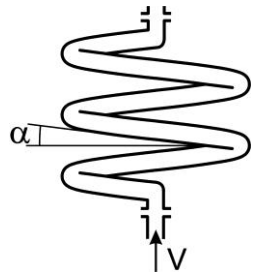

**9.21.** Невесомая труба согнута по винтовой линии радиусом *R* с углом наклона *α* и имеет *N* витков (см. рисунок). Начало и конец трубы выведены по радиусу на ось винтовой линии. Через подводы, обеспечивающие свободное вращение трубы вокруг оси винтовой линии, прокачивается вода со скоростью *V*. Найти угловую скорость вращения трубы.

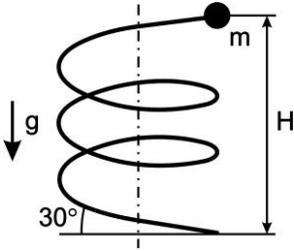

**9.22.** Тело массой *m* соскальзывает с высоты *H* по винтовому желобу с радиусом *R* и углом наклона 30° (см. рисунок). Желоб массой *M* может свободно вращаться вокруг своей вертикальной оси. Найти угловую скорость вращения желоба после соскальзывания тела. Трения нет.

**9.23.** Три одинаковые точечные массы соединены невесомыми стержнями длиной *L* и образуют равносторонний треугольник ABC, который вращается в своей плоскости с угловой скоростью Ω вокруг оси, проходящей через центр (см. рис. *a*). Связь AC исчезает. Найти линейные скорости масс в момент, когда они займут положение, показанное на рис. *b*.

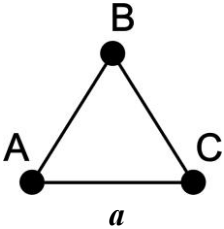
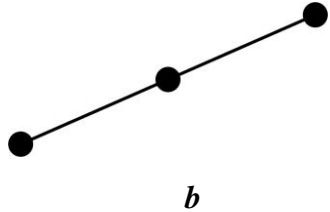

*a*                       *b*

**9.24**. На гладком столе лежит диск массой *M*, радиусом *R*, вращающийся с угловой скоростью $\omega_0$. На диск падает плашмя без горизонтальной скорости круглая пластилиновая лепешка массой *M*, радиусом *R*/2. Край лепешки совпадает с краем диска. Найти угловую скорость диска с приклеившейся к нему лепешкой.

**9.25**. Сплошной цилиндр радиусом *R*, вращающийся с угловой скоростью ω, ставят вертикально на шероховатую горизонтальную плоскость. Коэффициент трения μ. Сколько оборотов сделает цилиндр?

**9.26**. Два одинаковых диска, насаженных на гладкие оси, один из которых вращался с угловой скоростью ω, привели в соприкосновение (см. рисунок). Найти установившуюся скорость вращения дисков. Какая часть энергии перейдет в тепло?

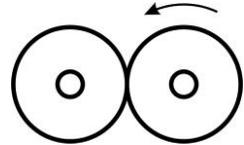

**9.27**. К диску, раскрученному вокруг оси, проходящей через центр симметрии перпендикулярно плоскости, поднесли два таких же неподвижных диска, так что они начали раскручиваться вокруг своих осей, тормозя первый диск. Какой будет угловая скорость диска к моменту, когда скорости дисков уравняются?

**9.28**. Цилиндрическая банка с жидкостью раскручена вокруг оси симметрии так, что жидкость не успела закрутиться. Как изменится угловая скорость вращения к моменту, когда угловые скорости банки и жидкости уравняются? Сколько энергии перейдет в тепло? Моменты инерции жидкости и пустой банки равны.

**9.29**. Диск массой $M$, радиусом $R$ может вращаться вокруг своей оси без трения. На тонкую шершавую ось, находящуюся на расстоянии $\rho$ от центра диска, насадили диск массой $m$, радиусом $r$, вращающийся с угловой скоростью $\omega_0$. Найти установившуюся угловую скорость вращения системы.

### 9.3. Физический маятник

**9.30**. Найти частоту малых колебаний тонкого стержня длиной $L$ в поле тяжести вокруг горизонтальной оси, перпендикулярной стержню и расположенной на расстоянии $x$ от его середины.

**9.31**. Симметричный крест, состоящий из двух взаимно перпендикулярных тонких однородных стержней длиной $L$, может колебаться в поле тяжести вокруг горизонтальной оси, проходящей через один из стержней и перпендикулярной ему. При каком удалении $X$ оси вращения от центра креста период его малых колебаний будет минимален? Найдите минимальное значение периода колебаний креста.

**9.32**. Обруч подвешен за верхнюю точку и может колебаться в вертикальной плоскости. Найти частоту малых колебаний: а) в плоскости обруча; б) перпендикулярно плоскости обруча.

**9.33**. Найти частоту малых колебаний тонкостенной сферы вокруг своей хорды в поле тяжести. Каким будет период малых колебаний для шара?

**9.34**. Тонкостенный сферический сосуд радиусом $R$ целиком заполнен водой и совершает малые колебания относительно точки подвеса О, удаленной на расстояние $L = 2R$ от центра сферы. Вода постепенно, слой за слоем, намерзает на внутреннюю поверхность сосуда. Во сколько раз изменится амплитуда колебаний к моменту, когда вся вода замерзнет? Время полного замерзания воды много больше периода колебаний.

**9.35**. Каркас массой *m,* сделанный из тонкой проволоки, имеет вид полуокружности с диаметром. Каркас шарнирно закреплён в средней точке диаметра и может колебаться перпендикулярно своей плоскости. Какова частота малых колебаний, если диаметр полуокружности *d*?

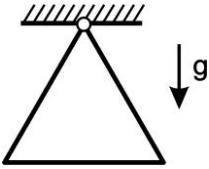

**9.36**. Найти частоту малых колебаний проволочного равностороннего треугольника, подвешенного на шарнире за вершину в поле тяжести (см. рисунок). Треугольник колеблется в плоскости рисунка.

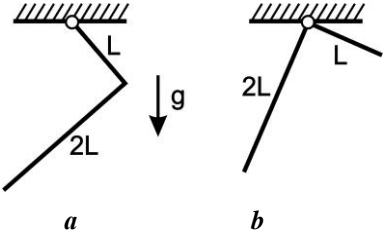

**9.37**. Однородный стержень согнут под прямым углом и подвешен на шарнире (см. рисунок). Найти частоту малых колебаний стержня для двух случаев, показанных на рис. *a* и *b*.

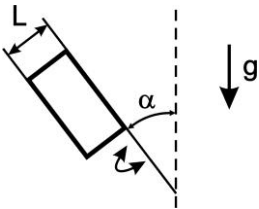

**9.38**. Найти частоту малых колебаний прямоугольной пластинки в поле тяжести относительно оси, которая проходит через край пластинки и наклонена под углом α к вертикали (см. рисунок).

**9.39**. Номерок из гардероба представляет собой диск радиусом *R*, на краю которого имеется отверстие радиусом *r*. Номерок висит на тонком гвозде. Найти частоту его малых колебаний в своей плоскости.

**9.40**. В ободе колеса, имеющего форму диска радиусом *R*, массой *M*, застрял камешек массой *m*. Найти частоту малых колебаний при плоском покачивании колеса без проскальзывания на горизонтальной плоскости.

**9.41**. Обруч радиусом *R* лежит горизонтально внутри неподвижной гладкой сферы радиусом 2*R*. Найти частоту малых колебаний обруча под действием силы тяжести.

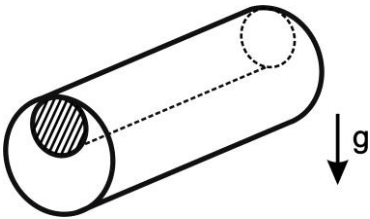

**9.42**. В цилиндре радиусом *R* параллельно оси на расстоянии *R/2* вырезан цилиндрический канал, в который вставлена цилиндрическая вставка радиусом *R/2* и длиной, равной длине цилиндра (см. рисунок). Плотность вещества цилиндра $\rho_1$, вставки $-$ $\rho_2 < \rho_1$, трение между стенками цилиндра и вставкой равно нулю. Найти частоту малых колебаний цилиндра в поле тяжести вокруг горизонтальной оси, совпадающей с осью большого цилиндра.

**9.43.** На краю шара радиусом $R$, плотностью $\rho_1$ вырезана сферическая полость радиусом $R/2$ (см. рисунок). В полость вставлена шаровая вставка плотностью $\rho_2 < \rho_1$ того же радиуса. Трения между стенками полости и вставкой нет. Найти частоту малых колебаний шара в поле тяжести

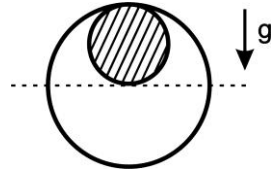

вокруг горизонтальной оси, совпадающей с диаметром большого шара.

**9.44.** Найти частоту малых колебаний полушара радиусом $R$ на горизонтальной плоскости в поле тяжести. Проскальзывания нет.

**9.45.** На неподвижный горизонтальный стержень радиусом $r$ надета тонкостенная труба радиусом $R$. Найти частоту малых колебаний трубы в поле тяжести. Труба движется без проскальзывания.

**9.46.** Найти частоту малых колебаний шарика радиусом $r$, двигающегося без проскальзывания по внутренней поверхности сферы радиусом $R$ в поле тяжести.

**9.47.** Найти частоту малых колебаний стержня массой $m$, длиной $L$, прикрепленного верхним концом к шарниру, а нижним – к середине нерастянутой горизонтальной пружины жесткостью $k$, концы которой закреплены.

**9.48.** Однородный стержень подвешен к потолку на двух одинаковых нитях, закрепленных на концах стержня. В положении равновесия нити вертикальны. Найти частоту малых крутильных колебаний стержня вокруг вертикальной оси, проходящей через центр стержня.

**9.49.** Найти частоту малых крутильных колебаний вокруг вертикальной оси для однородного кольца, подвешенного к потолку на трех одинаковых нитях. В положении равновесия нити вертикальны и делят кольцо на три равные части. Плоскость кольца горизонтальна.

**9.50.** На гладком столе лежат стержни AB и BC, соединенные в точке B шарниром. Точки A и C соединены невесомой пружиной жесткостью $k$ (см. рисунок). Длины $L$ и массы $M$ стержней одинаковы. В положении равновесия стержни образуют угол $60^0$. Найти частоту малых колебаний. Как изменится результат, если закрепить точку B?

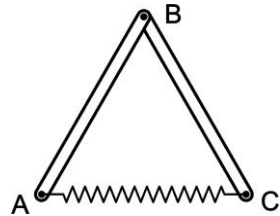

**9.51**. Диск массой *M,* радиусом *R* может катиться без скольжения по прямолинейному горизонтальному рельсу. К центру диска шарнирно прикреплен невесомый стержень длиною *L*, на конце которого находится точечная масса *m*. Найти период малых колебаний такого маятника.

### 9.4. Плоское движение тел

**9.52**. Описать движение гантели, состоящей из двух одинаковых шаров массой *m* и радиусом *r* каждый, соединенных невесомым стержнем длины *R* >> *r* (см. рисунок), после упругого лобового столкновения частицы массой 2*m*, скоростью *V* с одной из масс гантели. До столкновения гантель покоилась.

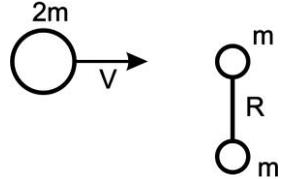

**9.53**. По стержню длиной *L*, лежащему на гладком столе, наносится удар. Направление удара перпендикулярно стержню. При ударе концы стержня приобретают скорости *V* и 2*V*. Найти расстояние от точки удара до середины стержня.

**9.54**. На льду лежит стержень длиной *L*, массой *M*, с которым упруго сталкивается шарик массой *m*. Скорость шарика *V* направлена по нормали к стержню. Точка удара близка к концу стержня. Найти скорость шарика после столкновения.

**9.55**. На гладком столе лежит стержень массой *M*, длиной *L*. Перпендикулярно стержню движется шарик массой *m*. После упругого удара о стержень шарик остановился. На каком расстоянии от середины стержня произошел удар?

**9.56**. Каким участком сабли следует рубить лозу, чтобы рука не чувствовала удара? Саблю считать однородным стержнем длиной *L*.

**9.57**. Однородный диск радиусом *R*, вращавшийся с угловой скоростью ω, разбился по диаметру на две равные части. Найти скорости поступательного и вращательного движения осколков.

**9.58.** Две одинаковые гантельки, движущиеся навстречу со скоростями *V* и 2*V* (см. рисунок), сталкиваются концами и слипаются. Описать движение образовавшейся гантельки. Длина каждой гантельки *L*.

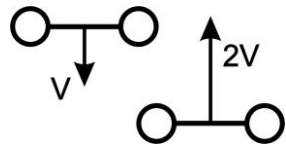

**9.59.** Две одинаковые гантельки, вращающиеся в одной плоскости навстречу друг другу с угловыми скоростями Ω и 2Ω (см. рисунок),

сталкиваются концами и слипаются. Описать движение образовавшейся гантельки.

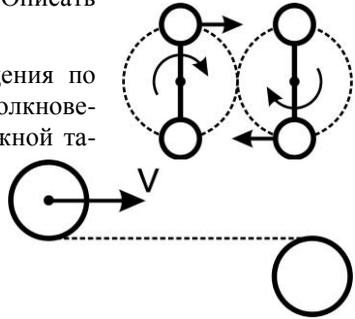

**9.60.** Шайба, двигавшаяся без вращения по льду со скоростью $V$, при касательном столкновении склеивается с первоначально неподвижной такой же шайбой (см. рисунок) и через некоторое время после совместного движения отрывается от нее. Какой будет максимальная скорость второй шайбы после отрыва?

**9.61.** На обруч массой $m$, радиусом $R$ намотана тонкая веревка линейной плотностью $\rho$. Обруч катится по плоскости, при этом веревка разматывается. Считая, что плоскость абсолютно шероховатая, найти зависимость скорости обруча от времени. Длина веревки $L$, начальная скорость обруча $V_0$.

**9.62.** В одном из хоккейных матчей шайба, летевшая без вращения со скоростью $V$ под углом $45^0$ к борту, упруго ударилась о борт и отскочила… по нормали. Найти скорость вращения отскочившей шайбы. Масса шайбы $m$, радиус $R$.

**9.63**. Какое максимальное число оборотов вокруг своей оси может сделать футбольный мяч после одиннадцатиметрового удара? Радиус мяча 0,11 м. Начальная скорость мяча направлена под малым углом к горизонту. Сопротивлением воздуха пренебречь.

**9.64.** На шероховатый пол падает шар радиусом $R$, который вращается с угловой скоростью $\omega$ по часовой стрелке вокруг горизонтальной оси и имеет горизонтальную скорость $V_0 > \omega R$, перпендикулярную оси вращения (см. рисунок). Найти угловую скорость вращения шара и его горизонтальную скорость после $N$ упругих отскоков от пола.

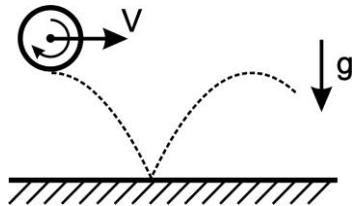

**9.65**. Стержень длиной $L$, движущийся поступательно со скоростью $V$, одним концом задевает за закрепленную стенку (см. рисунок). Найти угловую скорость вращения стержня после удара. Угол между осью стержня и начальной скоростью $\alpha$. Удар упругий. Рассмотреть два случая: а) трения нет; б) проскальзывания нет.

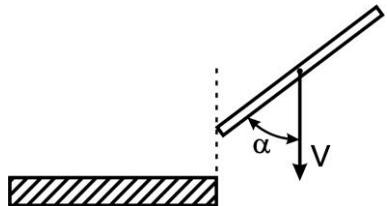

**9.66.** По льду скользят две шайбы, имевшие одинаковые начальные скорости. Одна из них при этом вращается, а другая движется только поступательно. Какая из шайб пройдет большее расстояние? Коэффициент трения не зависит от скорости.

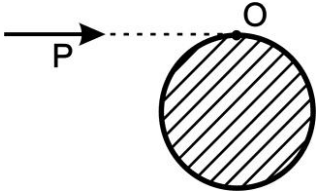

**9.67.** Лежащей на льду шайбе радиусом *R* касательным ударом в точку О сообщают импульс *P* (см. рисунок). Сколько оборотов сделает шайба и какое расстояние пройдет до остановки? Сила трения $F = -\alpha SV$ пропорциональна скорости *V* и площади контакта *S* шайбы со льдом.

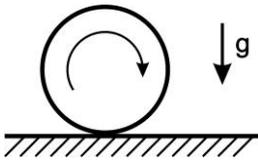

**9.68**. Однородный цилиндр раскрутили вокруг оси и поставили без поступательной скорости на шероховатую горизонтальную плоскость. Ось вращения параллельна плоскости (см. рисунок). Нарисовать синхронные графики зависимости от времени угловой скорости, поступательной скорости, ускорения и смещения. Какая доля энергии цилиндра перейдет в тепло?

**9.69**. Определить минимальное значение угловой скорости, при которой обруч, брошенный вперед с закруткой, сможет покатиться назад. Определить его установившуюся скорость, если начальная угловая скорость превышает минимальную.

**9.70**. Определить ускорение скатывания с наклонной плоскости с углом наклона к горизонту α: а) полого и сплошного цилиндров; б) шара.

**9.71**. Определить ускорение, с которым цилиндрическая бочка массой *m*, целиком заполненная жидкостью массой *M*, скатывается без проскальзывания с наклонной плоскости с углом наклона α. Вязкостью жидкости и моментом инерции днищ бочки пренебречь.

**9.72.** Шар радиусом *R*, массой *M* начинает скользить по наклонной плоскости с углом наклона α. Коэффициент трения скольжения μ. С какой

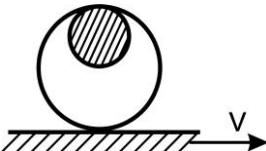

силой соскальзывающий шар действует на плоскость?

**9.73**. В цилиндре с радиусом основания *R* параллельно оси вырезан цилиндрический канал радиусом *R/2*. В канал вставлена цилиндрическая вставка из того же материала. Расстояние между осями главного цилиндра и вставки равно *R/2*, трения между цилиндром и вставкой нет. Цилиндр поставили на шероховатую плоскость, движущуюся со скоро-стью *V* (см. рисунок). Найти установившуюся угловую скорость цилиндра.

**9.74**. В цилиндре с радиусом основания *R* параллельно оси вырезан цилиндрический канал радиусом *R/2* (см. рисунок). В канал вставлена цилиндрическая вставка из того же материала. Расстояние между осями главного цилиндра и вставки равно *R/2*, трение между цилиндром и вставкой отсутствует. Найти угловое ускорение, с которым цилиндр будет скатываться без проскальзывания с наклонной плоскости.

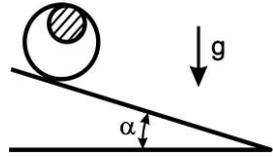

**9.75**. Однородный цилиндр раскрутили вокруг оси и поставили без начальной скорости на наклонную плоскость (см. рисунок). Описать движение цилиндра при различных соотношениях между коэффициентом трения и углом наклона.

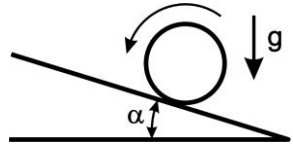

**9.76**. Найти время, за которое цилиндр радиусом *r*, вращающийся с начальной угловой скоростью ω, достигнет наивысшего положения, будучи поставлен на шероховатую наклонную плоскость с углом наклона к горизонту α.

**9.77**. На какую высоту поднимется поставленный на наклонную плоскость с углом наклона α вращающийся цилиндр радиусом *R*, энергией *T*, если имеется проскальзывание? Коэффициент трения $\mu$.

**9.78.** Рулон тонкой бумаги общей длиной *L*, намотанный на невесомую трубу радиусом *R*, раскручивается под действием силы тяжести по наклонной плоскости с углом наклона α к горизонту (см. рисунок). За какое время рулон развернется наполовину? Проскальзывания нет.

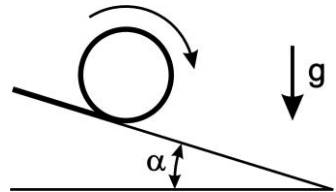

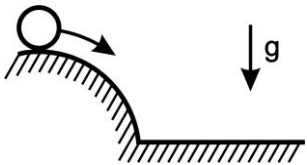

**9.79**. С верхней точки цилиндрической горки радиусом *R* скатывается без проскальзывания и без начальной скорости цилиндр радиусом *r* (см. рисунок). На какой высоте цилиндр оторвется от горки?

**9.80**. На шероховатом столе стоит палочка, которая начинает падать из вертикального положения в поле тяжести. При угле наклона палочки $45^0$ ее нижний конец начинает скользить. Найти коэффициент трения.

**9.81.** Палочка длиной *L*, массой *M,* стоящая вертикально на горизонтальной плоскости, начинает падать с нулевой начальной скоростью. Найти силу, с которой палочка действует на плоскость перед моментом полного касания в случаях: а) плоскость гладкая;  б) проскальзывания нет.

**9.82.** Лестница длиной *L*, опирающаяся верхним концом на гладкую вертикальную стену, а нижним – на гладкий горизонтальный пол, начинает падать. Опишите ее движение, если в начальный момент она покоилась, а расстояние от нижнего конца до стены было равно *d*. Определите силы реакции стены и пола при движении лестницы.

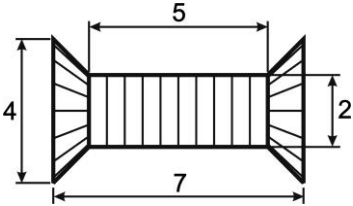

**9.83**. С каким ускорением надо тянуть вверх нить, намотанную на катушку, чтобы катушка не падала? Плотность материала катушки ρ. Размеры катушки указаны на рисунке.

**9.84**. Исследовать движение маятника Максвелла с моментом инерции *J*, весом *mg* и радиусом оси *r*. Найти зависимость натяжения нитей маятника *F* от времени.

### 9.5. Вращение с изменением ориентации оси.  Гироскоп

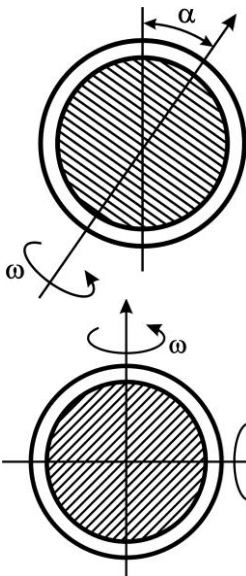

**9.85.** В укреплённую на вертикальной оси тонкую сферическую оболочку массой *M,* радиусом *R* вложен шар массой *M* и радиусом *R.* Шар раскручен до угловой скорости ω под углом α к вертикали (см. рисунок). Найти установившуюся угловую скорость вращения оболочки и шара, если между ними действует сила трения.

**9.86.** В тонкую сферическую оболочку массой *M*, радиусом *R* вложен шар массой *M,* радиусом *R*. Оболочка и шар раскручены до угловых скоростей ω в перпендикулярных направлениях (см. рисунок). Найти установившуюся угловую скорость вращения оболочки и шара, если между ними действует сила трения.

**9.87**. Найти частоту прецессии волчка, вращающегося с большой угловой скоростью вокруг своей оси, под действием силы тяжести.

**9.88**. Волчок, имеющий форму диска диаметром 30 см, насаженного посредине оси длиной 20 см, вращается с угловой скоростью 15 об/с вокруг оси симметрии (см. рисунок). Определить угловую скорость регулярной прецессии волчка.

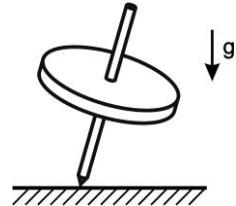

**9.89**. Исследовать устойчивость движения обруча, катящегося без проскальзывания с угловой скоростью ω по горизонтальной плоскости. Плоскость обруча вертикальна. Радиус обруча $r$.

**9.90**. Исследовать устойчивость движения однородного обруча, вращающегося вокруг вертикального диаметра с угловой скоростью ω. Нижней точкой обруч соприкасается с горизонтальной плоскостью. Радиус обруча $r$.

**9.91**. Исследовать устойчивость свободного вращения спичечного коробка вокруг оси, проходящей через его центр и параллельной одной из его сторон.

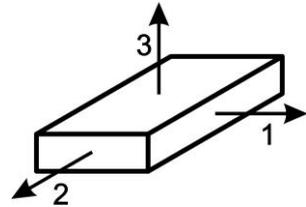

# 10. НЕИНЕРЦИАЛЬНЫЕ СИСТЕМЫ ОТСЧЕТА

**10.1.** Оценить силу давления воздуха на пол и потолок в свободно падающем лифте.

**10.2.** Оценить, с какой точностью выполняется первый закон Ньютона в системе отсчета спутника, движущегося по круговой орбите и повернутого к поверхности Земли одной и той же стороной.

**10.3.** Точка движется относительно диска по закону $r = r(t)$. Диск вращается относительно наблюдателя с постоянной угловой скоростью $\omega$. Найти закон движения точки, ее скорость и ускорение в системе отсчета неподвижного наблюдателя.

**10.4**. Тело свободно падает с высоты 500 м на землю. Принимая во внимание вращение Земли и пренебрегая сопротивлением воздуха, определить, насколько отклонится тело при падении. Географическая широта места $60^0$.

**10.5.** Вращение Земли вызывает наклон поверхности воды в реке. Оценить угол наклона поверхности воды для реки, текущей с севера на юг на широте $\varphi$.

**10.6.** На каком расстоянии от орудия упадет снаряд, выпущенный вертикально вверх со скоростью $V$ на широте $\varphi$, если пренебречь сопротивлением воздуха?

**10.7.** Оценить влияние вращения Земли на малые колебания математического маятника (маятник Фуко). Нарисовать проекции траектории на горизонтальную плоскость при различных начальных условиях. Какой будет частота прецессии маятника на экваторе, если угловая амплитуда его колебаний $\theta$?

**10.8.** Тонкая кольцевая стеклянная трубка заполнена водой и вращается в поле тяжести вокруг своего вертикального диаметра. В трубке имеется пузырек воздуха. При какой скорости вращения трубки равновесие пузырька в ее нижней точке будет устойчивым? Диаметр кольца $D$.

**10.9.** Тонкая U-образная трубка заполнена жидкостью и вращается вокруг своей оси симметрии (см. рисунок). Найти частоту колебаний столба жидкости в поле тяжести. Полная длина столба жидкости $L$. Капиллярными эффектами пренебречь.

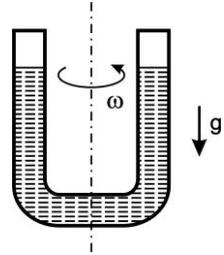

**10.10.** Оценить разницу между экваториальным и полярным радиусами Земли. Землю считать эллипсоидом вращения, гравитационный потенциал на поверхности которого меняется по закону

$$U = -\frac{GM}{r} + \frac{a_2}{2}\frac{GM}{r}\left[\frac{R^2}{r^2}\left(3\cos^2\theta - 1\right)\right],$$

где $M$ – масса Земли, $R$ – ее радиус на экваторе, $r, \theta$ – сферические координаты точки на поверхности эллипсоида, причем угол $\theta = 0$ соответствует северному полюсу. Численный коэффициент $a_2 = 1,1\cdot10^{-3}$.

**10.11.** Вертикальная U-образная трубка, заполненная водой, вращается вокруг одной из своих половин с угловой скоростью $\omega$. Расстояние между прямолинейными частями трубки $L$. Концы трубки открыты. Найти разность уровней воды в трубке. Оценить разность глубины океана на полюсе и на экваторе, обусловленную вращением Земли. В среднем глубина океана 3 км.

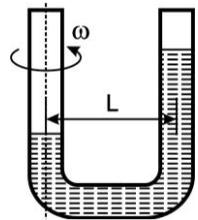

**10.12.** Трубка вращается с постоянной угловой скоростью вокруг вертикальной оси, составляя с ней угол $45^0$. В трубке находится тяжелый шарик. Определить движение шарика, если в начальный момент его скорость относительно трубки была равна нулю, а начальное расстояние от точки пересечения трубки с осью равнялось $L$.

**10.13**. Шарик массой *m*, прикрепленный к концу горизонтальной пружины жесткостью *k*, находится в положении равновесия в трубке на расстоянии *L* от вертикальной оси. Определить относительное движение шарика, если трубка, образующая с осью прямой угол, начинает вращаться вокруг вертикальной оси с постоянной угловой скоростью $\omega$.

**10.14**. Горизонтальная трубка длиной *L* равномерно вращается вокруг вертикальной оси с угловой скоростью $\omega$. Внутри трубки свободно скользит пробка. Определить скорость пробки относительно трубки в момент вылета и время движения пробки в трубке. В начальный момент пробка покоилась на расстоянии $x_0$ от оси.

**10.15**. В узкий канал, проходящий через центр Земли и экватор, опустили тело с нулевой начальной скоростью. Найти время падения тела до центра Земли и его скорость в центре. Угловая скорость вращения Земли $\Omega$, коэффициент трения о стенки канала $\mu$.

**10.16**. На палочку длиной *L* надета бусинка массой *m*. Коэффициент ее трения о палочку $\mu$. Палочка вращается по конусу с углом раствора $2\alpha$ с угловой скоростью $\omega$ относительно вертикальной оси, проходящей через конец. Пренебрегая весом бусинки, написать уравнение ее движения во вращающейся системе координат.

**10.17.** Стержень длиной *L*, массой *M* шарнирно подвешен за верхнюю точку в поле тяжести *g* и равномерно вращается с угловой скоростью $\omega$ вокруг вертикальной оси. При каком угле наклона стержня к вертикали может происходить такое вращение?

**10.18**. Тонкое гладкое проволочное кольцо радиусом *R* вращается с постоянной угловой скоростью $\omega$ вокруг своего диаметра (см. рисунок). На кольцо надета бусинка, которая в начальный момент находилась в точке А с угловой координатой $\theta_0 = 60^0$ и имела относительно кольца скорость $\omega \cdot R \cdot \sin\theta_0$. Найти время движения бусинки до точки В.

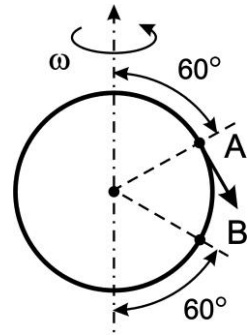

**10.19.** Оценить высоту прилива, вызываемого Солнцем и Луной.

**10.20.** Космическая станция представляет собой гантель длиной $d$, вращающуюся с угловой скоростью $\Omega$ вокруг вертикальной оси, проходящей через центр масс (см. рисунок). В отсеках станции, каждый из которых имеет массу $M$, есть резервуары, один из которых в начальный момент заполнен топливом массой $m$, а другой пуст. Какую работу необходимо затратить, чтобы полностью перекачать топливо из одного отсека в другой?

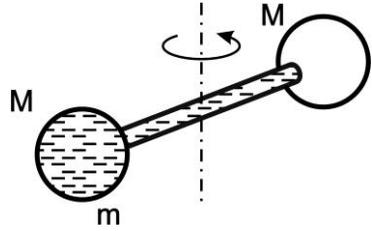

**10.21.** К Z-образной трубке (см. рисунок) через подвижный подвод того же сечения посредине трубки подается вода со скоростью $V$. Вода вытекает из обоих концов трубки, вызывая ее вращение. Найти угловую скорость вращения трубки, если ее длина $2L$.

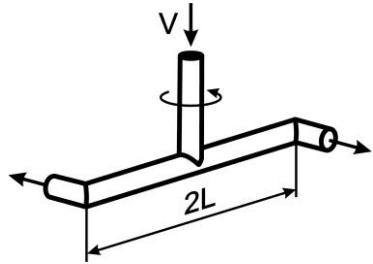

**10.22.** Найти закон движения самолета в атмосфере, при котором наиболее точно имитируется состояние невесомости.

# ЛИТЕРАТУРА